\documentclass[aps,pre,preprint,superscriptaddress,nofootinbib]{revtex4-1}
\usepackage[T1]{fontenc}
\usepackage[utf8]{inputenc}
\usepackage{color}
\usepackage{float}
\usepackage{amsmath}
\usepackage{amssymb}
\usepackage{graphicx}
\usepackage{hyperref}
\usepackage{appendix}
\usepackage{xcolor}
\usepackage{comment}

\makeatletter

\@ifundefined{showcaptionsetup}{}{
 \PassOptionsToPackage{caption=false}{subfig}}
\usepackage{subfig}
\makeatother

\begin{document}
\title{Role of bridge nodes in epidemic spreading: Different regimes and crossovers}

\author{Jing Ma}
\email{jingma@bu.edu}
\affiliation{Department of Physics,
  Boston University, 590 Commonwealth Ave., Boston, Massachusetts 02215, USA}

\author{Lucas D. Valdez}
\affiliation{Instituto de Investigaciones
  F\'isicas de Mar del Plata (IFIMAR), FCEyN,
  Universidad Nacional de Mar del Plata-CONICET, D\'ean Funes 3350,
  (7600) Mar del Plata, Argentina}
\affiliation{Department of Physics,
  Boston University, 590 Commonwealth Ave., Boston, Massachusetts 02215, USA}

\author{Lidia A. Braunstein}
\affiliation{Instituto de Investigaciones
  F\'isicas de Mar del Plata (IFIMAR), FCEyN,
  Universidad Nacional de Mar del Plata-CONICET, D\'ean Funes 3350,
  (7600) Mar del Plata, Argentina}
\affiliation{Department of Physics,
  Boston University, 590 Commonwealth Ave., Boston, Massachusetts 02215, USA}

\begin{abstract}

Power-law behaviors are common in many disciplines, especially in network science. Real-world networks, like disease spreading among people, are more likely to be interconnected communities, and show richer power-law behaviors than isolated networks.
In this paper, we look at the system of two communities which are connected by bridge links between a fraction $r$ of bridge nodes,
and study the effect of bridge nodes to the final state of the Susceptible-Infected-Recovered model, by mapping it to link percolation.
By keeping a fixed average connectivity, but allowing different transmissibilities along internal and bridge links,
we theoretically derive different power-law asymptotic behaviors of the total fraction of the recovered $R$ in the final state as $r$ goes to zero, for different combinations of internal and bridge link transmissibilities.
We also find crossover points where $R$ follows different power-law behaviors with $r$ on both sides when the internal transmissibility is below but close to its critical value, for different bridge link transmissibilities.
All of these power-law behaviors can be explained through different mechanisms of how finite clusters in each community are connected into the giant component of the whole system, and enable us to pick effective epidemic strategies and to better predict their impacts.

\end{abstract}

\maketitle


\section{Introduction}

Network theory is a powerful tool that can be applied in many disciplines.
In this framework, real systems such as the power grid, the brain, and societies are represented by a network \cite{barabasi2016network},
which is a graph composed of nodes and links that represent the interaction between nodes.
Many researchers use network theory to study the spreading of an epidemic in order to predict its evolution and to implement strategies to decrease
its impact in healthy populations \cite{newman2002spread}. Diseases like Ebola \cite{faye2015chains}, H1N1 \cite{eastwood2010responses}, and the novel coronavirus COVID-19 \cite{world2020coronavirus} spread not only domestically, but also from one country to another, mainly through air transportation \cite{gardner2013global}. These international airports are bridge nodes, which establish connections between more than one community.
In this work, we explore how bridge nodes affect the disease spreading.

The most used model that reproduces the final state of nonrecurrent epidemics is the
Susceptible-Infected-Recovered (SIR) model \cite{bailey1975mathematical,anderson1992infectious,newman2002spread}. In this model a
susceptible individual (S) in contact with an infected one (I) gets
infected with probability $q$ at each time step. An infected individual recovers (R) after $t_r$ time steps since it was infected. Once an
individual is recovered, it does not play any role in the spreading. In
this model the fraction of recovered individuals $R$ is the order
parameter of a continuous phase transition with a control parameter
$T=1-(1-q)^{t_r}$, where $T$ is the effective probability of
infection denoted as the transmissibility.
It is known that there exists a critical value $T_c$ that separates a nonepidemic phase from an epidemic phase, so that in the thermodynamic limit $R=0$ for $T\le T_c$, and $R>0$ for $T>T_c$ \cite{cohen2010complex,newman2010networks}.
It was shown \cite{grassberger1983critical,newman2010networks,stauffer2018introduction} that the final state of the SIR model can be mapped into link percolation, due to the fact that infecting through a link in SIR is equivalent to
occupying a link in link percolation,
and thus the final state of SIR can be solved using percolation tools.
Each realization of the final state of SIR is one cluster in link percolation, and an epidemic corresponds to the giant component (GC) in link percolation, which is distinguished from outbreaks (corresponding to the finite clusters) by a threshold for the cluster size $s_c$ \cite{lagorio2009effects}. 
In random complex networks it is worthwhile to find exact solutions
for the main magnitudes of the final state of the SIR model using the
generating function formalism.  In this approach two generating
functions are used \cite{newman2001random,callaway2000network}.  One of them is the generating function of the degree distribution $G_0(x)=\sum_k P(k) x^k$, where $P(k)$ is the degree distribution with $k_{\min} \leq k \leq k_{\max}$, and $k_{\min}$ and $k_{\max}$ are
the minimum and maximum degree respectively. The other is the generating function of the excess degree distribution $G_1(x)=\sum_k k P(k)/\langle k \rangle x^{k-1}$, where $\langle k \rangle$ is the average degree of the network.
In the SIR model for \emph{isolated} networks, the probability $f_{\infty}$ that a branch of infected nodes reach the infinity for a given transmissibility $T$ satisfies the self-consistent equation
$f_{\infty} =
1-G_1(1-T f_{\infty})$ \cite{braunstein2007optimal,newman2010networks}.
Note that $G_1(1-T f_{\infty})$ is the probability that following a random chosen
link, which leads to a node, the branch of infection does not reach the infinity through its $(k-1)$ outgoing links.
The fraction of recovered individuals $R$, which is equivalent to the fraction $P_{\infty}$ of nodes belonging to the GC in link percolation, is given by $R=
1-G_0(1-T f_{\infty})$ \cite{braunstein2007optimal,newman2010networks}, since $G_0(1-T f_{\infty})$ is the probability that a
random chosen node can not reach the infinity with infected nodes through any of its $k$ links. The critical value of the transmissibility is $T_c=1/(\kappa-1)$, where $\kappa=\langle k^2
\rangle/\langle k \rangle$ is the branching factor, and $\langle k^2
\rangle$ is the second moment of the degree distribution \cite{lagorio2011quarantine,buono2014epidemics}.
For Erd\"os-R\'enyi (ER) networks \cite{erdos1959random}, the degree follows a Poisson distribution $P(k)=\langle k \rangle^k e^{-\langle k \rangle}/k!\,$, and thus $T_c=1/\langle k \rangle$.
Around criticality $T_c$, many physical quantities behave as power laws, e.g., $P(s) \sim s^{-\tau+1}\exp(-s/s_{\max})$, where $P(s)$ is the probability to find a cluster of size $s$, $s_{\max}\sim |T-T_c|^{-1/\sigma}$ is the largest finite cluster size, and the fraction of recovered $R\sim |T-T_c|^{\beta}$ \cite{newman2010networks,cohen2002percolation}.

Before the last decade, researchers concentrated on studying these processes in isolated networks \cite{lagorio2011quarantine,pastor2001epidemic}.
However, real networks are rarely isolated \cite{gao2012networks,kenett2014network}. For example, each country has its own transportation network, and those networks from different countries are connected into a larger network due to international transportation. Also, different communities of people can hold different opinions, but their opinions can exchange through influencers. Thus it is more realistic to consider systems composed of many networks, which are called a network of networks (NON) \cite{buldyrev2010catastrophic,gao2011robustness,kenett2014network,gao2013percolation,kryven2019bond,kivela2014multilayer}.
A case of NON is a system composed of several communities (or layers), where a fraction of nodes $r$ from each layer are bridge nodes which are connected to bridge nodes from other communities through $k^b$ bridge links.
Bridge nodes, which can represent airports connecting countries, may have a huge impact on the system because they can influence individuals in other communities.
As these kinds of nodes are few compared to the number of nodes inside a community, it is reasonable to study these problems in the limit $r\rightarrow 0$.

In Ref.~\cite{dong2018resilience}, 
the authors studied node percolation in two ER communities with an ER distribution of bridge links, and studied the behavior of $R$ with $r$, in the limit $r\rightarrow 0$, with the constraint $r\langle k^b \rangle=\text{constant}$.
They found, using scaling relations, that $R \propto r^{1/\epsilon}$, where $r$ was associated to an external field.\footnote[2]{They used $\delta$ instead of $\epsilon$ since this behavior is analogous to the relation $M \propto H^{1/\delta}$ between the magnetization $M$ and the external field $H$ in the Ising model \cite{reynolds1977ghost}.}
In Ref.~\cite{valdez2018role}, the authors extended this result to an SIR model and also studied the dynamics. In the final state for $r\rightarrow 0$, they found the same value of the exponent as in Ref.~\cite{dong2018resilience}, and explained it from a geometrical point of view. In their interpretation, the GC was formed by finite clusters in both communities connected through bridges links at $T=T_c(r=0)$. Thus the exponent $\epsilon$ was associated with the exponent $\tau$ of the finite cluster size distribution, which allowed them to derive this exponent theoretically and obtained $\epsilon=1/(\tau-2)$.
Note that due to the constraint $r\langle k^b \rangle = \text{constant}$, the average external connectivity $\langle k^b \rangle$ diverges as $r\rightarrow 0$. In addition, they used the same transmissibility $T$ along intra- and interlinks.

However, from a realistic point of view, the fraction of bridge nodes and
the average external connectivity do not have to be related, and building a large number of connections for one node is practically expensive,
so it is unrealistic to study the case when $\langle k^b \rangle \rightarrow \infty$.
On the other hand, the interaction mechanisms are in general different for internal links than those for bridge links, and strategies like cutting international flights can be used to reduce the disease spreading, so the transmissibility along internal links and bridge links can be very different. In this paper, we use a more realistic approach in which the fraction of
bridge nodes $r$ and the average external connectivity $\langle k^b
\rangle$ are independent, and the transmissibility along bridges links
$T^b$ is different from the internal transmissiblity $T^I$.  When $r$ is small, we find very rich behaviors of $R$ with $r$, many of which are power laws $R\propto r^{1/\epsilon}$, depending on the values of $T^b$ and $T^I$. In these regions the exponent $\epsilon$ follows different functions of the exponents in the finite cluster size distributions.
Our theoretical results are in very good agreement with simulations. 


\section{Model}
In our model we consider a system composed of two communities $A$ and $B$,
with degree distributions $P^A(k)$ and $P^B(k)$. 
The communities are connected through a fraction $r$ of bridges nodes with degree distribution $P^b(k)$.
The transmissibility within each community is $T^I$ and the transmissibility along bridge links is $T^b$.
To reduce the number of parameters, we will assume that both communities have the same degree distribution, i.e., $P^A(k)=P^B(k) \equiv P(k)$.

Using the generation function formalism, the self-consistent equations of the system are given by
\begin{eqnarray}
\label{eq_f}
f &=& (1-r)\left[1-G_1(1-T^I f)\right] + r\left[1-G_1(1-T^I f)G_0^b(1-T^bf^b)\right],\\
\label{eq_fb}
f^b &=& 1-G_1^b(1-T^bf^b)G_0(1-T^I f),
\end{eqnarray}
where $f$ is the probability to expand a branch to the infinity through an internal link, $f^b$ is the probability to expand a branch to  the infinity through a bridge link, and $G_0(\cdot)$, $G_1(\cdot)$, $G_0^b(\cdot)$, and $G_1^b(\cdot)$ are the generating functions of the degree and excess degree distributions for internal and bridge links, respectively. The
first term on the RHS of Eq.~(\ref{eq_f}) is the contribution of nonbridge nodes that
transmit only internally, while the second term is the contribution of
bridge nodes which transmit both internally with $T^I$ and to the other
community through bridge links with $T^b$. Thus the fraction of recovered nodes of the system and the fraction of recovered nodes of bridge nodes are given by
\begin{eqnarray}
\label{eq_r}  R & = &(1-r)\left[1-G_0(1-T^I  f)\right] + r\left[1-G_0(1-T^I f)G_0^b(1-T^bf^b)\right], \\
\label{eq_rb}  R^b &= &\left[1-G_0(1-T^I f)G_0^b(1-T^bf^b)\right].
\end{eqnarray}

\begin{figure}[h!]
\begin{minipage}[t]{0.49\linewidth}\centering
\includegraphics[width=7cm]{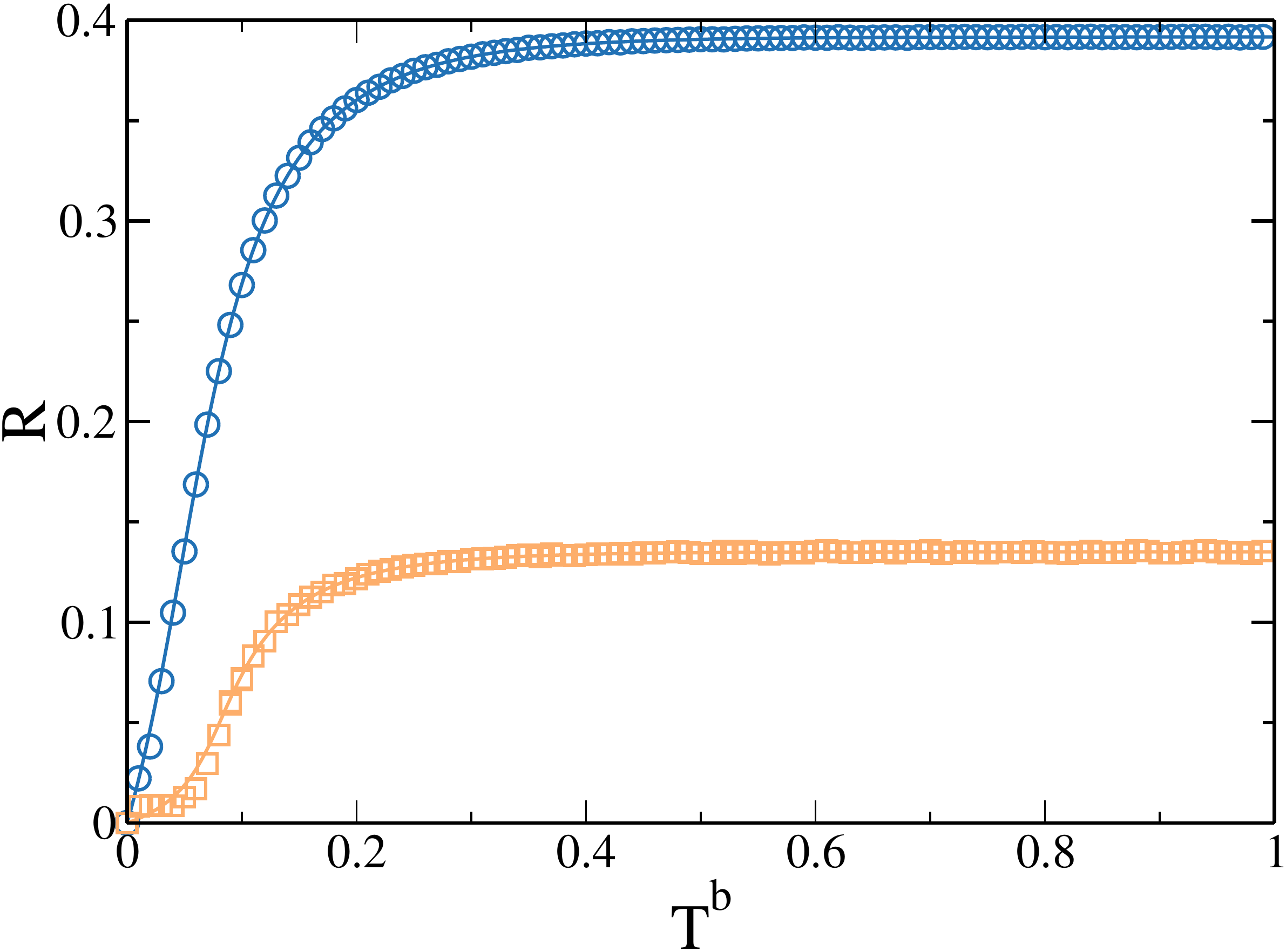}
\medskip
\centerline{(a)}
\end{minipage}\hfill
\begin{minipage}[t]{0.49\linewidth}\centering
\includegraphics[width=7cm]{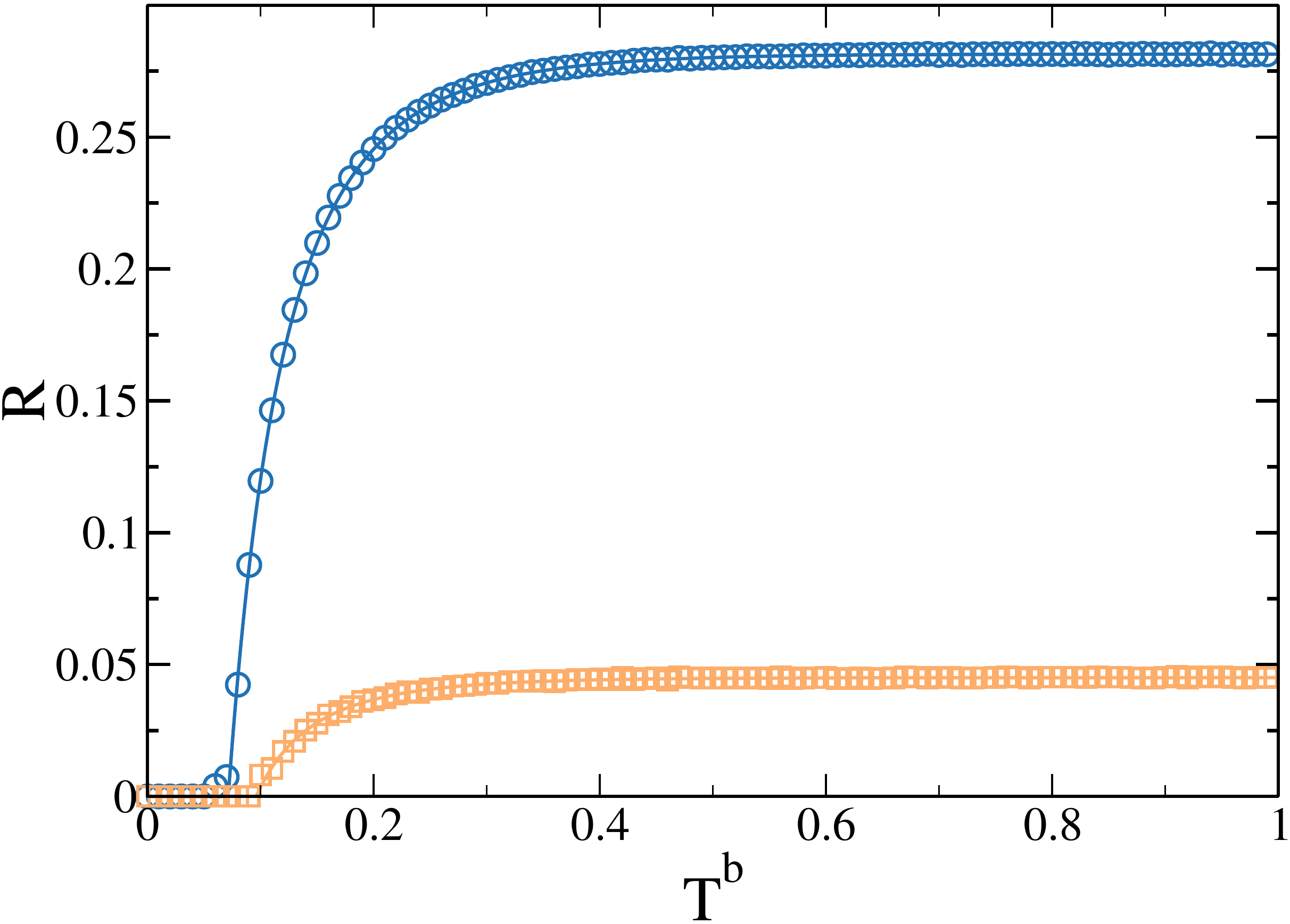}
\medskip
\centerline{(b)}
\end{minipage}
\caption{$R$ as a function of $T^b$ when both internal links and bridge links are ER networks, with $\langle k \rangle=4$ and $\langle k^b \rangle=10$, respectively. Theoretical solutions (dark blue solid lines for $r=0.1$ and light orange solid lines for $r=0.01$) are compared with stochastic simulation results of SIR in the final state (dark blue circles for $r=0.1$ and light orange squares for $r=0.01$), for (a) $T^I=0.25$ and (b) $T^I=0.2$. For the simulations, system sizes $N_A=N_B=10^5$, $k_{\min}=0$, $k_{\max}=100$, $s_c=200$, and are averaged over $10^3$ realizations.}
\label{fig_rvstb}
\end{figure}

In Fig.~\ref{fig_rvstb} we show the fraction of the recovered $R$ as a
function of $T^b$ for different values of $T^I$ and $r$, for a system where both internal links and bridge links follow an ER degree
distribution, with $\langle k \rangle=4$, and $\langle k^b \rangle=10$.
The solid lines show the numerical solutions obtained from Eqs.~(\ref{eq_f})-(\ref{eq_r}), and the square and circle symbols are the results from the SIR stochastic simulations. We can see that the theory agrees very well with the simulation results, and thus we will mainly use theoretical solutions hereafter.
Both theoretical solutions and simulation results in Fig.~\ref{fig_rvstb} show a critical value of $T^b$ that depends on $T^I$ and $r$.
The system is in a nonepidemic phase, with $R=0$, when $T^b \le T^b_c$, and is in an epidemic phase with a finite positive $R$ when $T^b > T^b_c$.
This is due to the fact that the self-consistent
Eqs.~(\ref{eq_f}) and (\ref{eq_fb}) have only one solution $f=f^b=0$ when $T^b \le T^b_c$, and a nontrivial physical solution exists only when $T^b>T^b_c$.

The theoretical value of $T^b_c$ can be obtained by solving
$|J-I|_{f,f^b=0}=0$, where $|\cdot|$ is the determinant, $J$ is the Jacobian matrix, and $I$ is the identity. Note that the elements of the
Jacobian matrix are given by $J_{i,j}= \left.\frac{\partial f_i}{\partial f_j}\right|_{f,f^b=0}$, where each of $f_i$ and $f_j$ represents $f$ or $f^b$. Thus explicitly $|J-I|_{f,f^b=0}=0$ can be written as
\begin{equation}\label{eq_critical}
    \begin{vmatrix}
      T^I(\kappa-1)-1 & rT^{b}_{c}\langle k^b\rangle \\[1ex]
      T^I\langle k \rangle & T^{b}_{c}(\kappa^b-1)-1
    \end{vmatrix}
    = 0.
\end{equation}
So $T^{b}_{c}$ is given by
\begin{equation}
\label{eq_tbc}
T^b_c =
\frac{T^I(\kappa-1)-1}{(T^I\left(\kappa-1\right)-1)(\kappa^b-1)-rT^I\langle
  k\rangle\langle k^b\rangle}.
\end{equation}

In the equation above, $T^b_c$ has physical meaning only when $T^I \le 1/(\kappa-1)$ (see Appendix \ref{app_sec_critical} for details).
This implies that any strategy that reduces the transmissibility between communities will prevent a macroscopic number of infected nodes only if the internal transmissibility is below the critical value for an isolated community.
We can see from Eq.~(\ref{eq_tbc}) that, as $r$ approaches $0$, the critical value $T^b_c(r \rightarrow 0)= 1/(\kappa^b-1)$.


\section{Different Regimes: Asymptotic Behaviors}

\begin{figure}[h!]
\centering
\includegraphics[width=10cm]{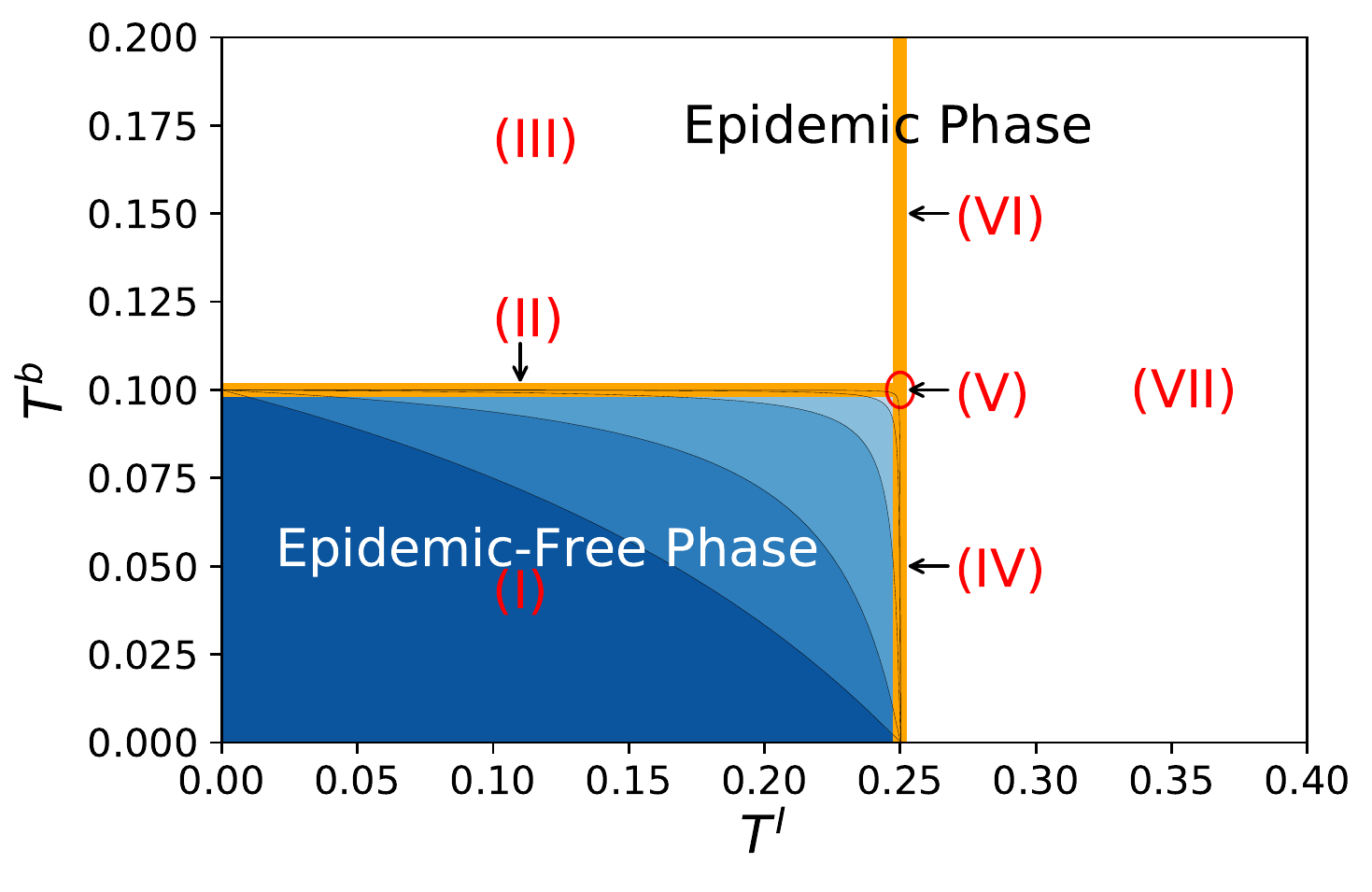}
\caption{Phase diagram for two ER communities connected by ER bridge links, with $\langle k \rangle=4$ and $\langle k^b \rangle=10$, respectively. The blue areas are the nonepidemic phases ($R=0$) for $r=0.5,0.1,0.01,0.001$, and the white area is the epidemic phase with $R>0$. As $r\rightarrow 0$, the nonepidemic phase expands and tends to be a rectangle.}
\label{fig_regimes}
\end{figure}

In Fig.~\ref{fig_regimes} we show the phase diagram for two ER communities connected by ER bridge links with $\langle k \rangle=4$ and $\langle k^b \rangle=10$, for different values of $r$.
As $r\rightarrow 0$, the nonepidemic phase tends to be a rectangle.
The boundaries of the rectangle are
$T^I = T^I_c=1/(\kappa-1)$ and $T^b=T^b_c(r\rightarrow 0)=1/(\kappa^b-1)$, which split the whole space into several regimes, where the relation between $R$ and $r$ follows different behaviors asymptotically.
In this section, we derive the asymptotic behavior of $R$ versus $r$, i.e., as $r\rightarrow 0$, by first looking at how $R$ depends on $(rR^b)$, and then how $R^b$ depends on $R$.

\begin{figure}[h!]
\begin{minipage}[t]{0.49\linewidth}\centering
\includegraphics[width=6cm]{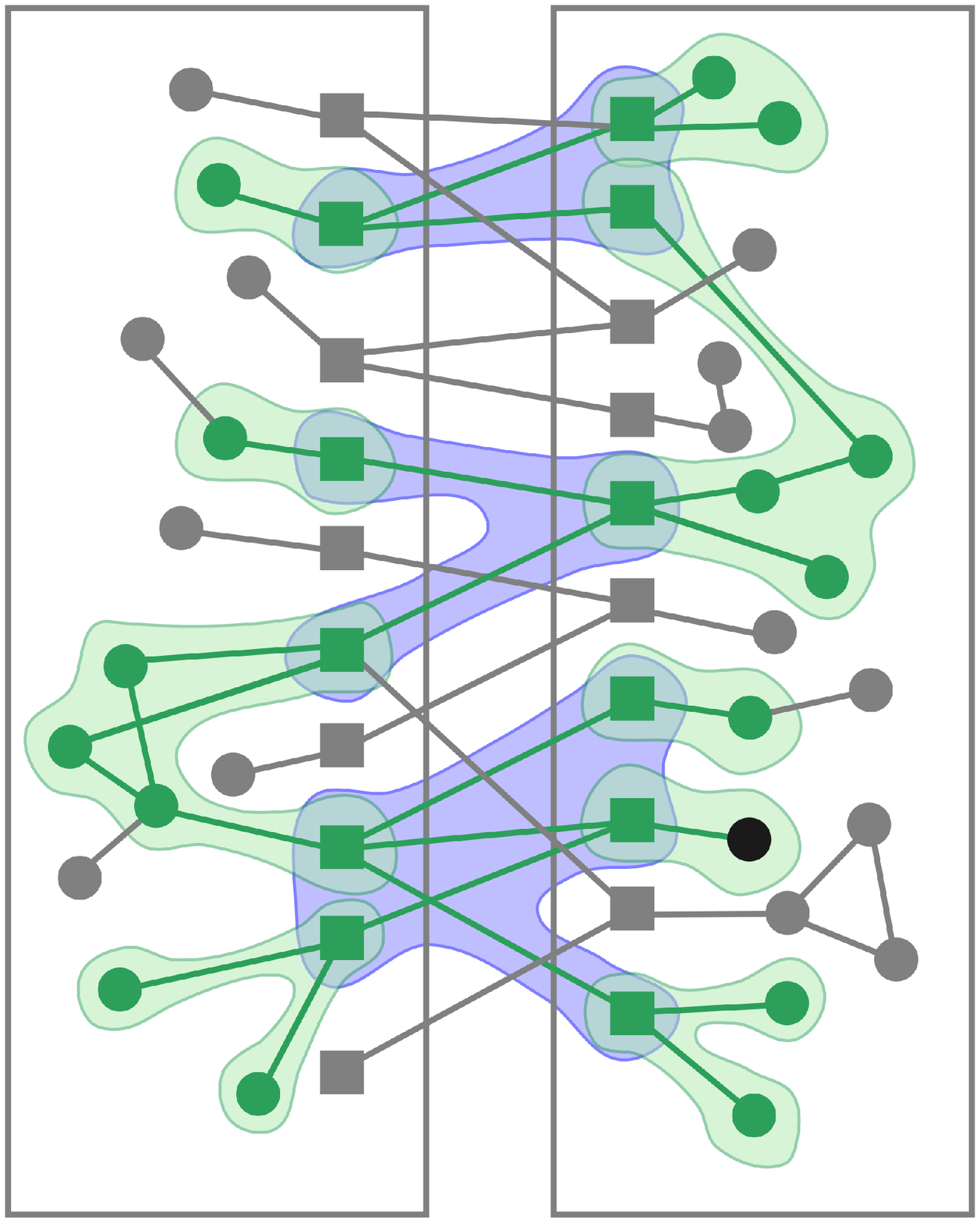}
\medskip
\centerline{(a)}
\end{minipage}\hfill
\begin{minipage}[t]{0.49\linewidth}\centering
\includegraphics[width=6cm]{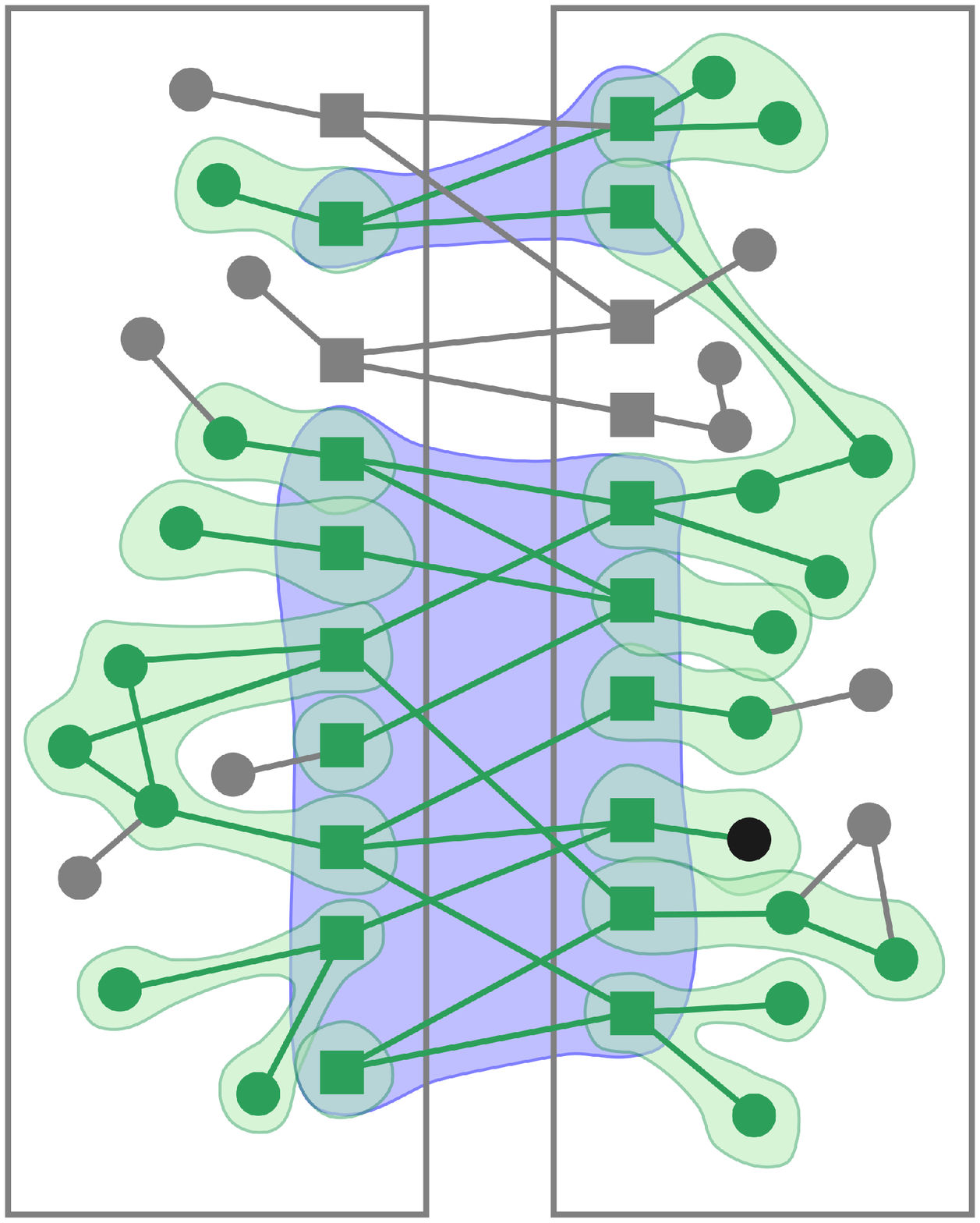}
\medskip
\centerline{(b)}
\end{minipage}
\caption{An illustration of how finite clusters of recovered nodes in each community (circled in green) are connected due to clusters by bridge links (circled in blue), and thus there exists a GC of recovered nodes in the entire system. (a) Only finite clusters exist in each community and for bridge links. (b) Only finite clusters exist in each community but a giant component exists for bridge links. All recovered nodes are plotted in green, except for patient zero, which is plotted in black, and all nodes that are never infected are plotted in gray. Links through which the disease is transmitted are plotted in green, while links that fail to transmit the disease are plotted in gray. Squares denote bridge nodes, and circles denote internal nodes.}
\label{fig_topology}
\end{figure}

When $T^I\le T^I_c$, there are only finite clusters of recovered nodes within each community.
However, in the epidemic phase, these finite clusters in each community are connected due to bridge links,
and thus form a GC of recovered nodes in the entire system, as illustrated in Fig.~\ref{fig_topology}.
Using the mapping between the SIR model and link percolation,
as any node in each community has a probability $r$ to be a bridge node, and each
bridge node has a probability $R^b$ to be recovered, a finite cluster
of size $s$ has a probability
$1-(1-r R^b)^s$ to have at least one recovered bridge node, and thus
belong to the GC of recovered.
Thus,
the size of the GC as $r \rightarrow 0$ is given by
\begin{equation}
\label{eq_rsum}
R = 1-\sum_{s=1}^\infty P(s)(1-r R^b)^s,
\end{equation}
where $P(s) \sim s^{-\tau+1}\exp(-s/s_{\max})$ is the probability of a finite cluster of size $s$ within a community, $\tau$ is the Fisher exponent of each community, and the largest finite cluster size $s_{\max}\sim |T^I-T^I_c|^{-1/\sigma}$. Then we can derive the behavior of $R$ with $r$ for $T^I$ below, or equal to the critical internal transmissibility.

At the critical value $T^I=T^I_c=1/(\kappa-1)$ so that $s_{\max}$ diverges and thus $P(s) \sim s^{-\tau+1}$, Eq.~(\ref{eq_rsum}) can be simplified into $R\propto (rR^b)^{\tau-2}$ [see Eqs.~(\ref{app_eq_r})~and~(\ref{app_eq_rat}) in Appendix \ref{app_sec_exponents} for details].
This is due to the fact that the average number of infected bridge nodes in each finite cluster of a community depends on the topology of the community, and thus depends on $\tau$.
When $T^I < T^I_c$, Eq.~(\ref{eq_rsum}) can be reduced to $R\propto rR^b$, due to the finite $s_{\max}$ [see Eqs.~(\ref{app_eq_r})~and~(\ref{app_eq_rbelow}) in Appendix \ref{app_sec_exponents} for details].
This is intuitive since as $T^I$ is so small that each finite cluster of a community has very few bridge nodes, then the number of nodes in the GC will be proportional to the number of bridge nodes in the GC [as in Fig.~\ref{fig_topology} (a)].

When $\kappa^b<\infty$, which is always the case in reality, we need to explore the behavior of $R^b$ as well.
For each cluster connected through bridge links,
each bridge node has a probability $1-G_0(1-T^If)$ to be connected to the GC through internal links.
So as $r \rightarrow 0$, a finite cluster of bridge nodes of size $s$ has a probability $1-[G_0(1-T^If)]^s$ to belong to the recovered bridge nodes, and thus
\begin{equation}
\label{eq_rbsum}
R^b = 1-\sum_{s}^\infty P^b(s)[G_0(1-T^If)]^s,
\end{equation}
where $P^b(s) \sim s^{-\tau^b+1}\exp(-s/s^b_{\max})$ is the probability of a finite cluster of size $s$ connected by bridge links, $\tau^b$ is the Fisher exponent of bridge links, and $s^b_{\max}\sim |T^b-T^b_c(r\rightarrow 0)|^{-1/\sigma^b}$ is the largest finite cluster size of bridge links.
From Eq.~(\ref{eq_r}) we know that $R\approx 1-G_0(1-T^If)$ as $r\rightarrow 0$, so
\begin{equation}
\label{eq_rbsumapprox}
R^b \approx 1-\sum_{s}^\infty P^b(s)(1-R)^s.
\end{equation}

At the critical value $T^b=T^b_c(r\rightarrow 0)=1/(\kappa^b-1)$,
$s^b_{\max}$ diverges, and thus $R^b\propto R^{\tau^b-2}$ [see Eq.~(\ref{app_eq_rbat}) in Appendix \ref{app_sec_exponents} for details].
When $T^b < T^b_c(r\rightarrow 0)$,
$R^b\propto R$ since $s^b_{\max}<\infty$ [see Eq.~(\ref{app_eq_rbbelow}) in Appendix \ref{app_sec_exponents} for details].
When $T^b>T^b_c(r\rightarrow 0)$,
most bridge nodes are connected into one big cluster through bridge links, so Eqs.~(\ref{eq_rbsum})~and~(\ref{eq_rbsumapprox}) do not apply and
$R^b$ is not a power law of $R$ [see Fig.~\ref{fig_topology} (b)].

In summary,
\begin{equation}
\label{eq_rvsrrb}
R \propto
\begin{cases}
rR^b, & \text{if } T^I<1/(\kappa-1) \\
(rR^b)^{\tau-2}, & \text{if } T^I=1/(\kappa-1) \\
\text{not a power law of } (rR^b), & \text{if } T^I>1/(\kappa-1)
\end{cases}
,
\end{equation}
\begin{equation}
\label{eq_rbvsr}
R^b \propto
\begin{cases}
R, & \text{if } T^b<1/(\kappa^b-1) \\
R^{\tau^b-2}, & \text{if } T^b=1/(\kappa^b-1) \\
\text{not a power law of } R, & \text{if } T^b>1/(\kappa^b-1)
\end{cases}
.
\end{equation}

\begin{table}[htb]
\centering
\begin{tabular}{c c c c c c c} 
\hline
\hline
  &\hspace*{5ex}& $T^b<\frac{1}{\kappa^b-1}$ &\hspace*{5ex}& $T^b=\frac{1}{\kappa^b-1}$ &\hspace*{5ex}& $T^b>\frac{1}{\kappa^b-1}$ \\ [1ex]
\hline
$T^I<\frac{1}{\kappa-1}$ && \begin{tabular}{@{}c@{}}$\varnothing$\vspace{-1ex} \\ (Regime I) \end{tabular} && \begin{tabular}{@{}c@{}}$\epsilon=1-(\tau^b-2)$\vspace{-1ex} \\ (Regime II) \end{tabular} && \begin{tabular}{@{}c@{}}$\epsilon=1$\vspace{-1ex} \\ (Regime III) \end{tabular}\\ [4ex]
$T^I=\frac{1}{\kappa-1}$ &&
\begin{tabular}{@{}c@{}}$\epsilon=\frac{1}{\tau-2}-1$\vspace{-1ex} \\ (Regime IV) \end{tabular} &&
\begin{tabular}{@{}c@{}}$\epsilon=\frac{1}{\tau-2}-(\tau^b-2)$\vspace{-1ex} \\ (Regime V) \end{tabular} && 
\begin{tabular}{@{}c@{}}$\epsilon=\frac{1}{\tau-2}$\vspace{-1ex} \\ (Regime VI) \end{tabular} \\ [4ex]
$T^I>\frac{1}{\kappa-1}$ && \begin{tabular}{@{}c@{}}$\varnothing$\vspace{-1ex} \\ (Regime VII) \end{tabular} && \begin{tabular}{@{}c@{}}$\varnothing$\vspace{-1ex} \\ (Regime VII) \end{tabular} && \begin{tabular}{@{}c@{}}$\varnothing$\vspace{-1ex} \\ (Regime VII) \end{tabular} \\ [1ex]
\hline
\hline
\end{tabular}
\caption{Asymptotic power-law behaviors of $R$ with $r$ in different regimes. The exponent $\epsilon$ in $R\propto r^{1/\epsilon}$ is independent of the specific values of $\kappa$ or $\kappa^b$, but varies with the regimes where the combination of $T^I$ and $T^b$ falls in. $\varnothing$ means there is no power-law relation in that regime.}
\label{tb_exponents}
\end{table}

Combining Eqs.~(\ref{eq_rvsrrb})~and~(\ref{eq_rbvsr}), we obtain the asymptotic power-law behaviors of $R$ with $r$ in many regimes. Different values of $\epsilon$ in the relation $R \propto r^{1/\epsilon}$ are summarized in Table~\ref{tb_exponents}.
As an example,
when both communities and the bridge links are all ER networks, we have $\tau = \tau^b = 5/2$, and in the limit $\kappa^b \rightarrow \infty$ so that $T^b>1/(\kappa^b-1)$ all the time, we obtain the same exponents that were found in Refs.~\cite{dong2018resilience,valdez2018role},
in which $r \langle k^b \rangle = \text{constant}$,
and $r \rightarrow 0$.
Note that the results in Table~\ref{tb_exponents} apply to networks with any degree distributions,
i.e., either internal or bridge links or both can be homogeneous or heterogeneous.
Also,
a similar methodology can be applied to a system when the two communities have different degree distributions,
i.e., $P^A(k)\ne P^B(k)$,
and we can still correctly predict the asymptotic power-law relations between $R$ and $r$ for all regimes (see Appendix \ref{app_sec_panepb} for details).

\begin{figure}[h!]
\begin{minipage}[t]{0.33\linewidth}\centering
\includegraphics[width=5cm]{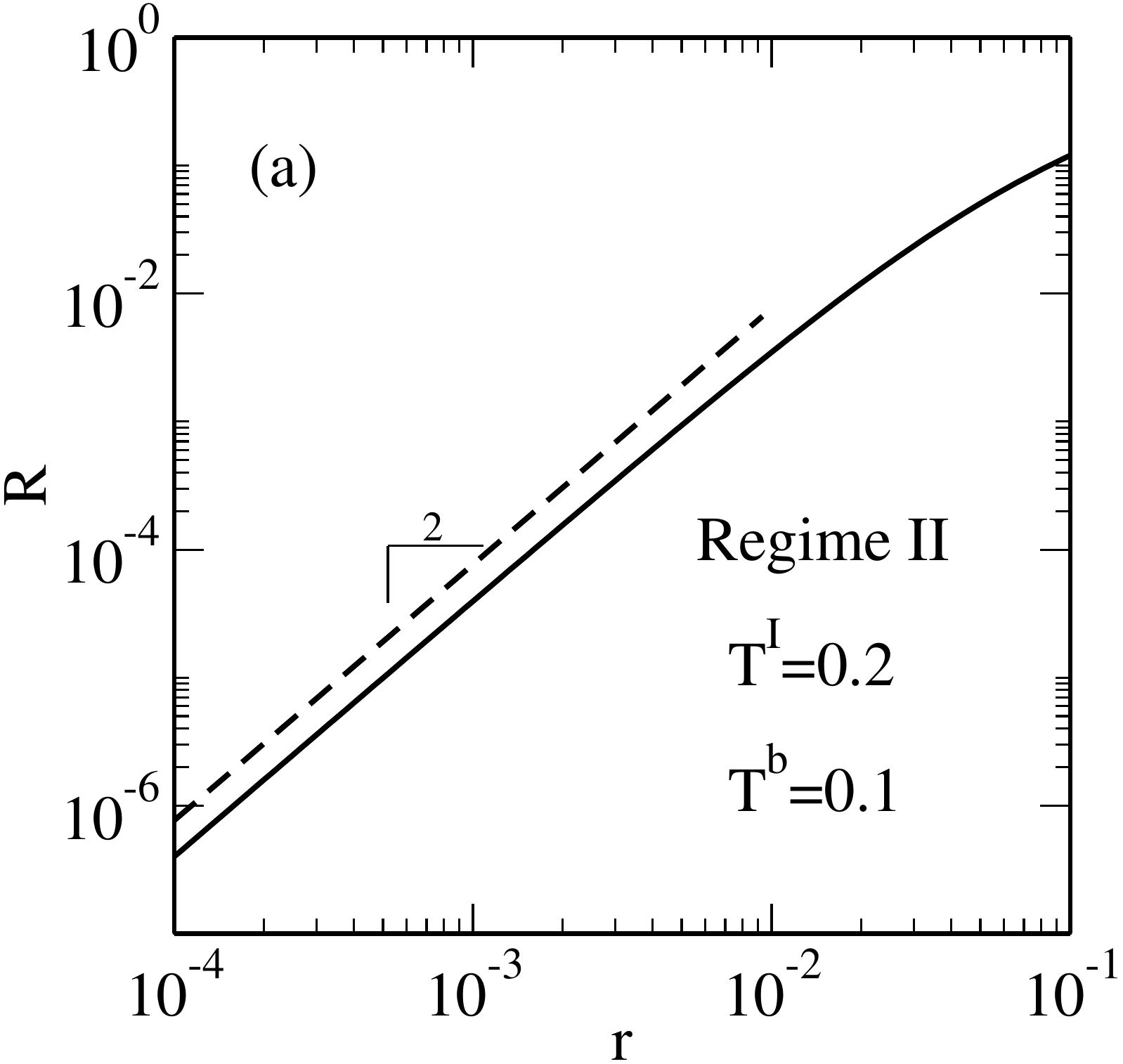}
\medskip
\end{minipage}\hfill
\begin{minipage}[t]{0.33\linewidth}\centering
\includegraphics[width=5cm]{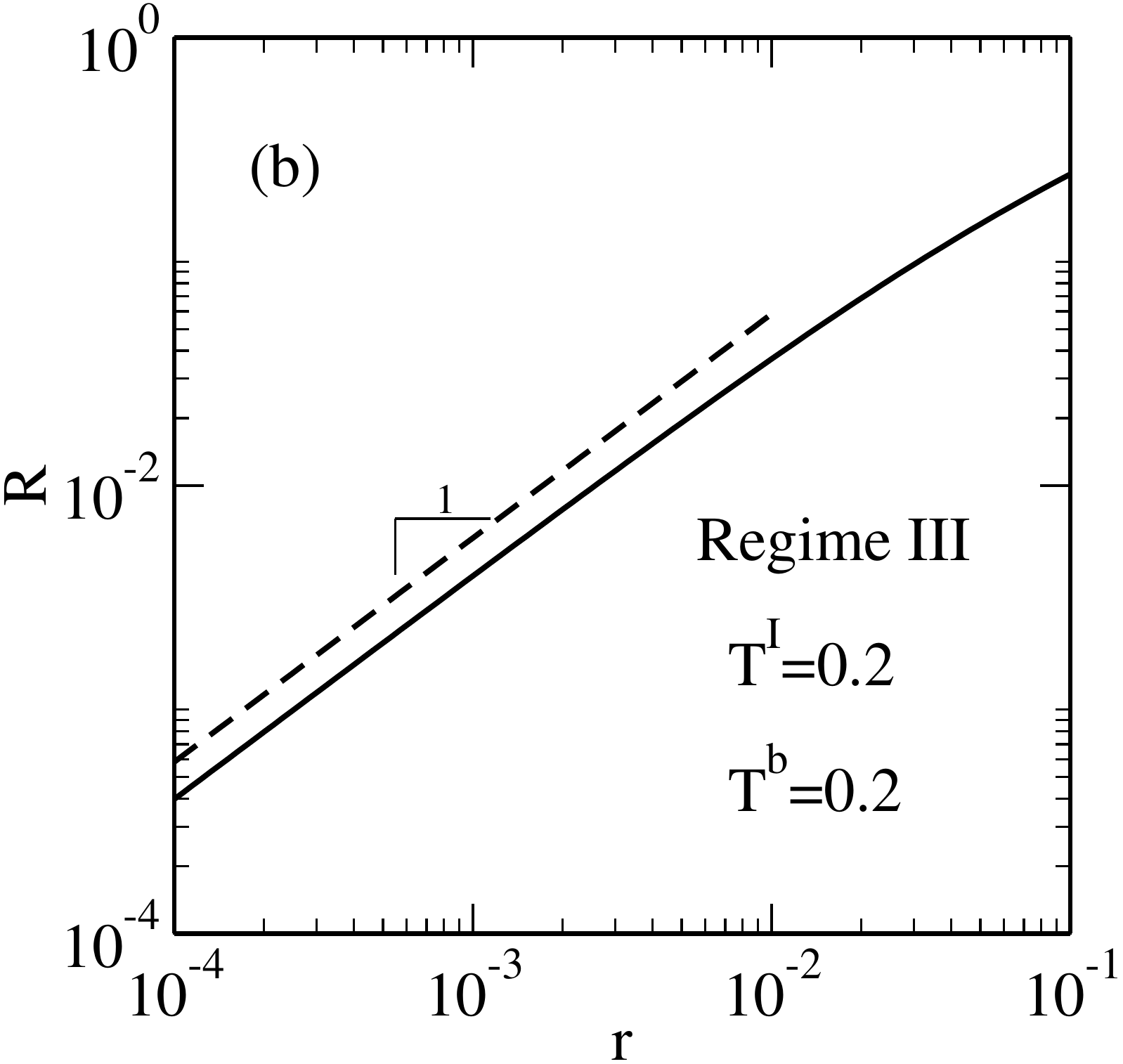}
\medskip
\end{minipage}\hfill
\begin{minipage}[t]{0.33\linewidth}\centering
\includegraphics[width=5cm]{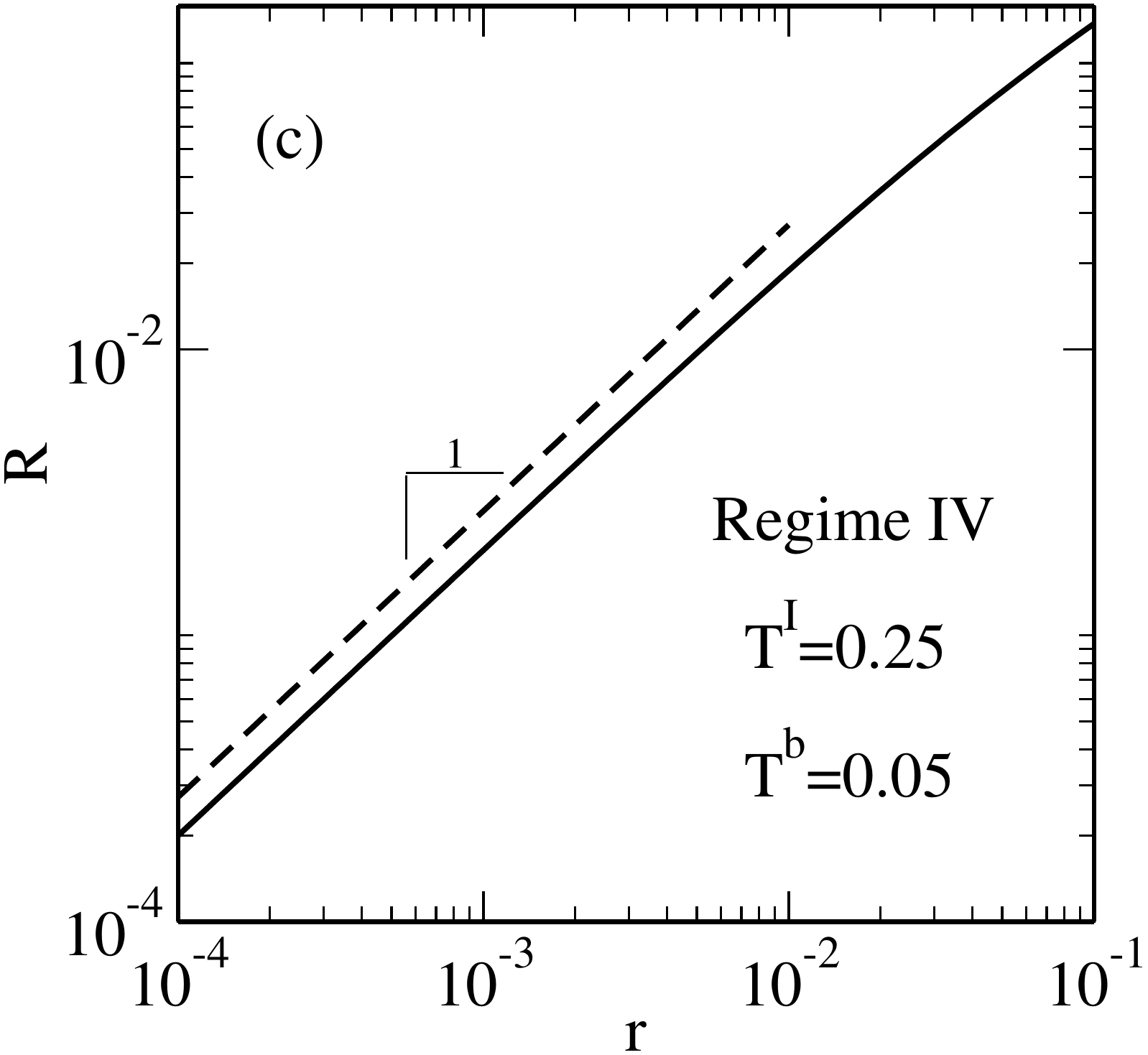}
\medskip
\end{minipage}\hfill
\vspace{1.5\baselineskip}
\begin{minipage}[t]{0.49\linewidth}\centering
\includegraphics[width=5cm]{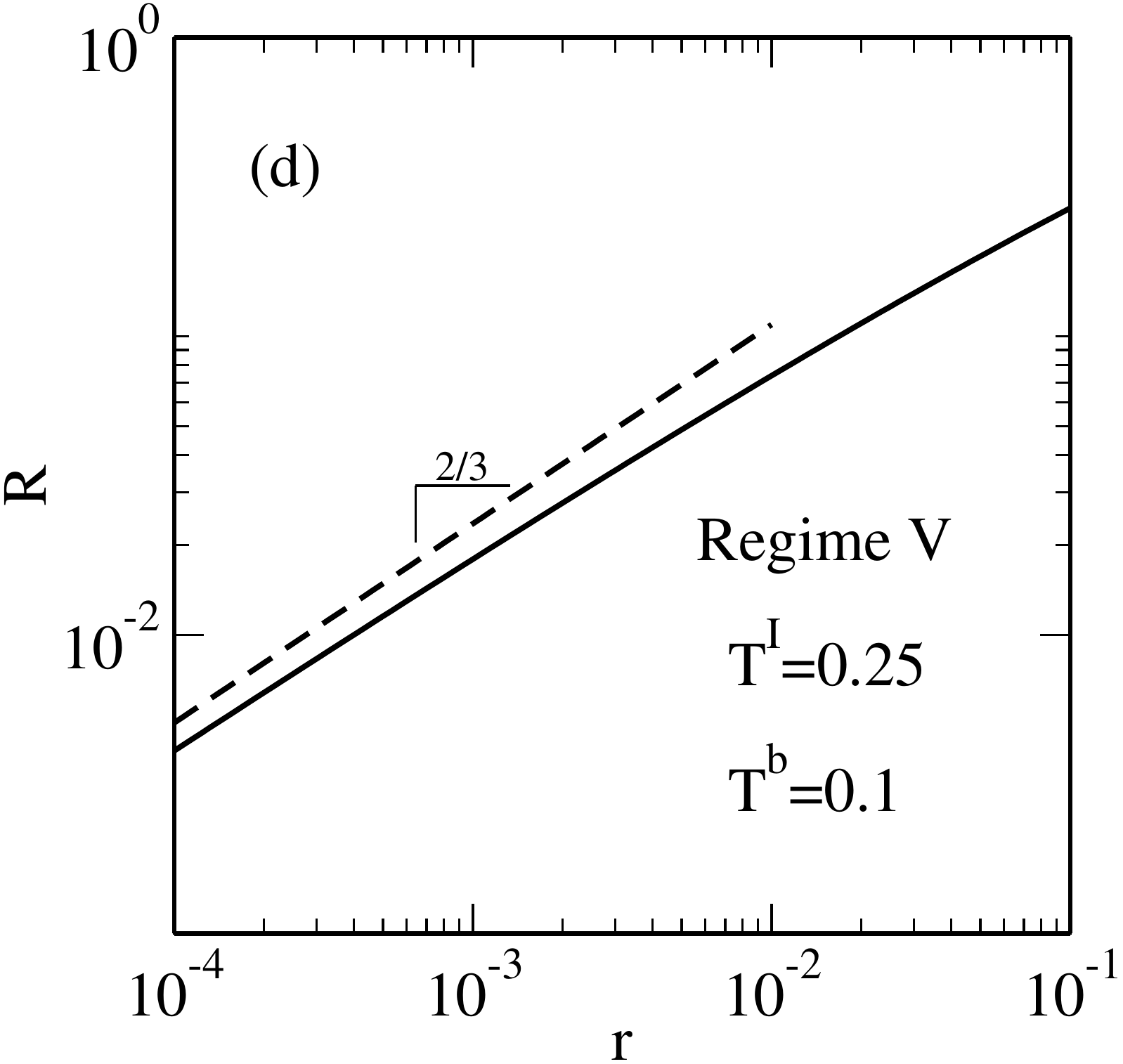}
\medskip
\end{minipage}\hfill
\begin{minipage}[t]{0.49\linewidth}\centering
\includegraphics[width=5cm]{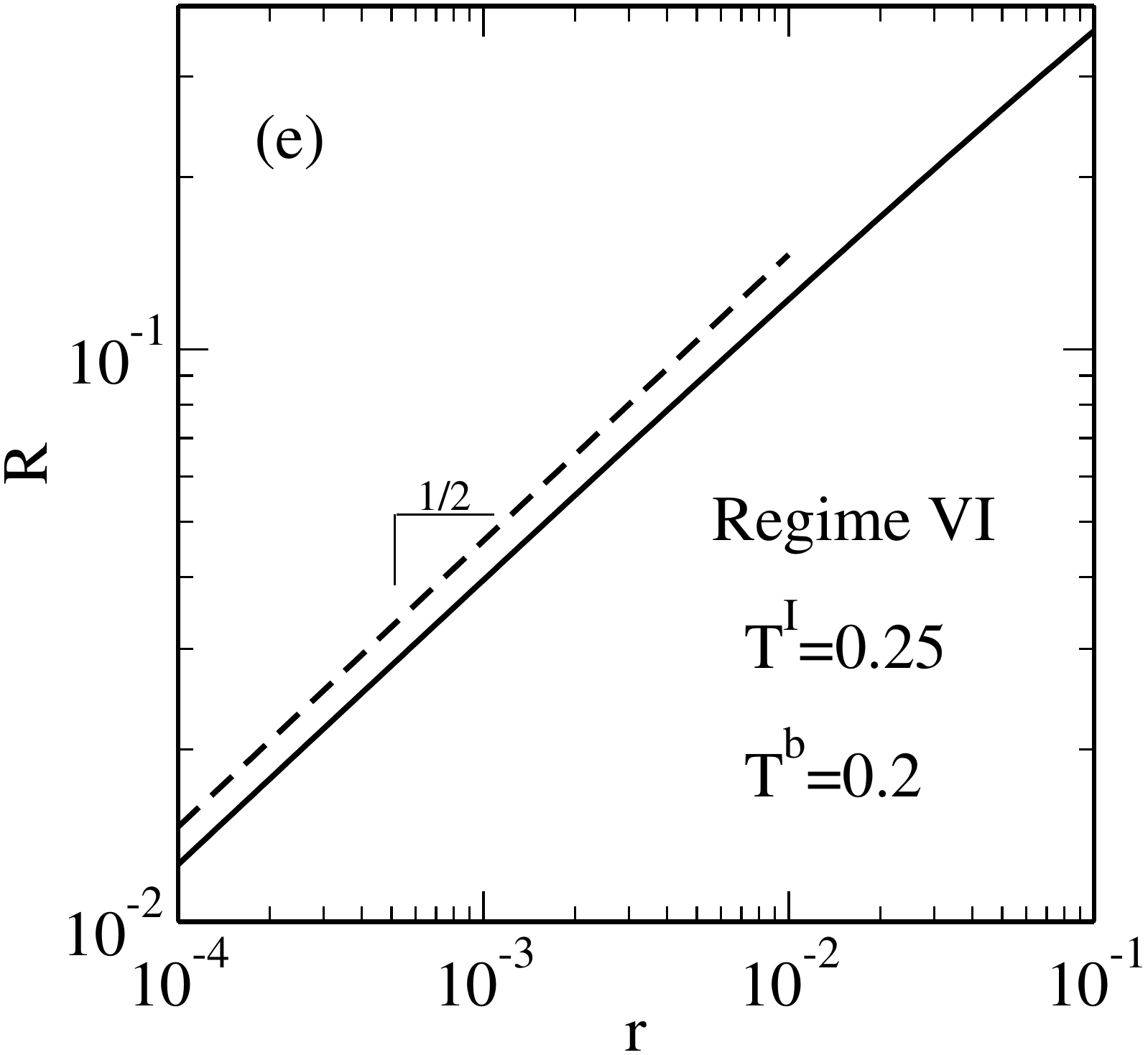}
\medskip
\end{minipage}
\caption{$R$ as a function of $r$ for different regimes where a power law exists: (a) Regime II: $T^I=0.2$, $T^b=0.1$; (b) Regime III: $T^I=0.2$, $T^b=0.2$; (c) Regime IV: $T^I=0.25$, $T^b=0.05$; (d) Regime V: $T^I=0.25$, $T^b=0.1$; and (e) Regime VI: $T^I=0.25$, $T^b=0.2$. Both internal links and bridge links are ER networks, with $\langle k \rangle=4$ and $\langle k^b \rangle=10$, respectively. In each regime, numerical solutions of Eqs.~(\ref{eq_f})-(\ref{eq_r}) are plotted in black solid lines, and a dashed line is drawn with the slope predicted by Table~\ref{tb_exponents}.}
\label{fig_exponents}
\end{figure}

In Fig.~\ref{fig_exponents} we show the numerical solutions of Eqs.~(\ref{eq_f})-(\ref{eq_r}) with the log-log plot of $R$ with $r$ for small $r$ in different regimes, for a system of two ER communities connected by ER bridge links, and thus $\tau=\tau^b=5/2$.
In each regime, we plot a dashed line with the slope predicted by the theory (see Table~\ref{tb_exponents}). We can see that our predictions are in good agreement with the numerical results. In Appendix \ref{app_sec_compare}, we also compared simulations using link percolation with numerical solutions. We use link percolation mapping instead of SIR to simulate the final state because the former is much less time-consuming for big system sizes. The simulation results agree well with theoretical solutions, except for some finite-size effects when $r$ is very small (see Appendix \ref{app_sec_compare} for details).

When a highly infectious epidemic occurs, one of the first strategies used by many countries is to shut down some international airports. Those international airports serve
as bridge nodes in the whole system of global transportation, so shutting them down is essentially reducing the percentage of bridge nodes $r$.
Meanwhile, international flights are cut for those airports that are still open, which mitigates the disease spreading by reducing the transmissibility along bridge links.
Also, social distancing strategies like staying at home as long as it is possible or wearing facial masks if having to go outside reduce the chance of face-to-face infection, which is utilized by most countries as another strategy to reduce both $T^I$ and $T^b$.

As can easily be seen from our results in this section, strategies like shutting down international airports are not as effective in some regimes as in others.
In those regimes with a smaller $\epsilon$, shutting down international airports to reduce $r$ will significantly reduce the fraction of recovered $R$, while in regimes with a larger $\epsilon$, $R$ will be reduced only slightly.
This helps us to decide what kind of strategies we are supposed to use to control disease spreading effectively; i.e., shutting down international airports had better be combined with strategies to reduce $T^I$ and $T^b$, so that it falls in a regime with a small $\epsilon$.

\section{Crossovers when $T^I\lesssim T^I_c$}

In Table \ref{tb_exponents}, we can see that the asymptotic values of the exponent $\epsilon$ change abruptly between regimes. However, in this section, we are going to show that, for $T^I \lesssim T^I_c$, the system behaves in the same way as $T^I=T^I_c$  for a relatively large value of $r$, but changes to its asymptotic behavior continuously as $r$ decreases.

From a percolation point of view, a finite cluster belongs to the GC if it contains recovered bridge nodes (with an overall percentage of $r R^b$).
As $r R^b$ decreases from $1$, the GC starts to lose some finite clusters so that $R$ also decreases from $1$.
When $r$ is not too small, finite clusters of \emph{smaller sizes} are more likely to be detached from the GC.
Note that the probability of a cluster of size $s$ is $P(s) \sim s^{-\tau+1}\exp(-s/s_{\max})$ for $T^I<T^I_c$ and $P(s) \sim s^{-\tau+1}$ for $T^I=T^I_c$, which are the same for smaller cluster sizes, so the behaviors of $R$ versus $r$ for different values of $T^I \lesssim T^I_c$ are the same when $r$ is not small enough.
The distribution $P(s)$ starts to differ when the GC starts to lose relatively large finite clusters, i.e., when $s$ is comparable to $s_{\max}$, and a crossover is going to show up.
Denote $r^*$ as where the crossover occurs, $R^*$ and $R^{b*}$ as the fraction of recovered nodes and bridge nodes at the crossover respectively, we have $r^* R^{b*}\sim 1/s_{\max}$ (see Appendix \ref{app_sec_exponents}).

\begin{figure}[h!]
\begin{minipage}[t]{0.33\linewidth}\centering
\includegraphics[width=5cm]{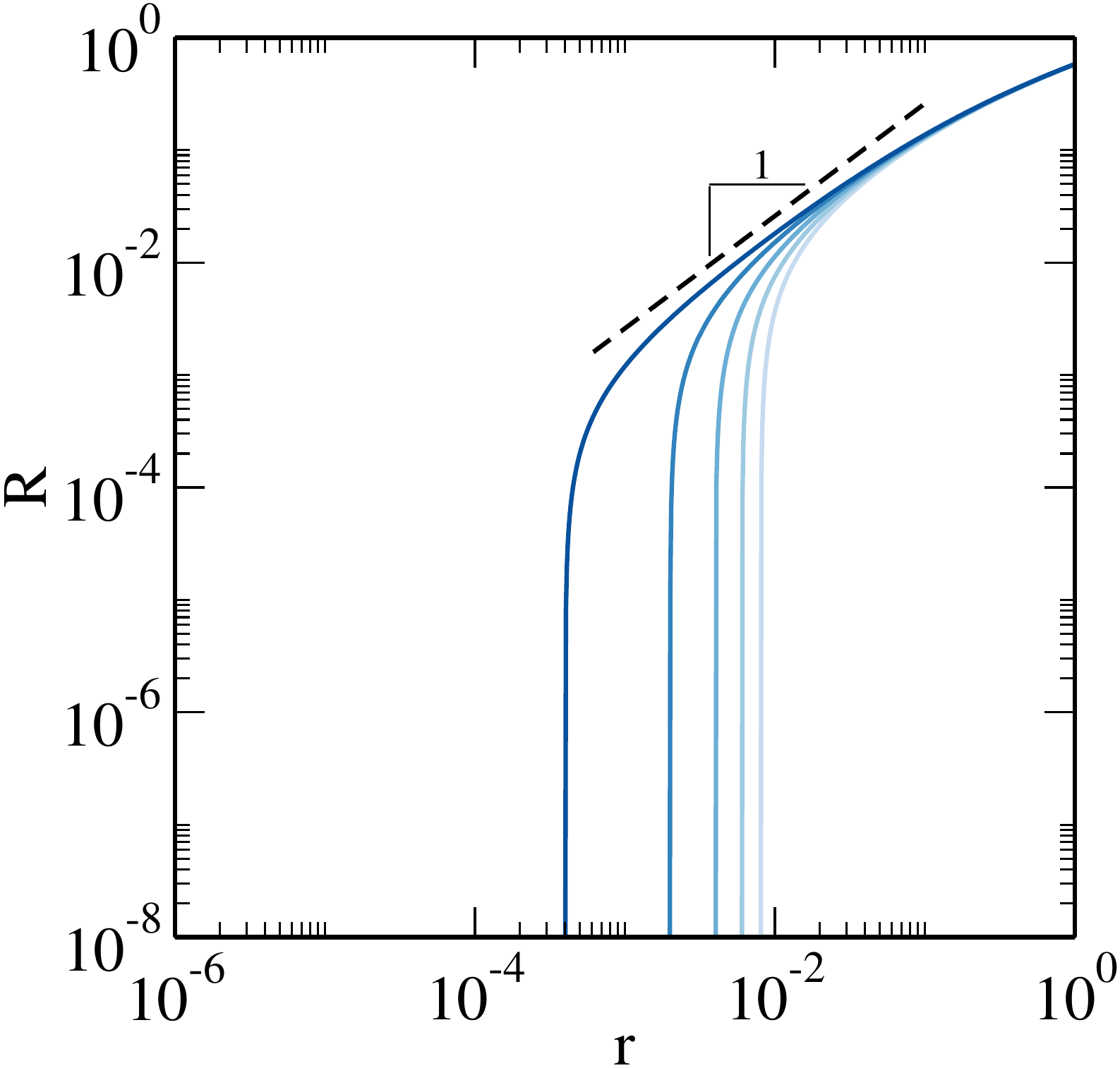}
\medskip
\centerline{(a)}
\end{minipage}\hfill
\begin{minipage}[t]{0.33\linewidth}\centering
\includegraphics[width=5cm]{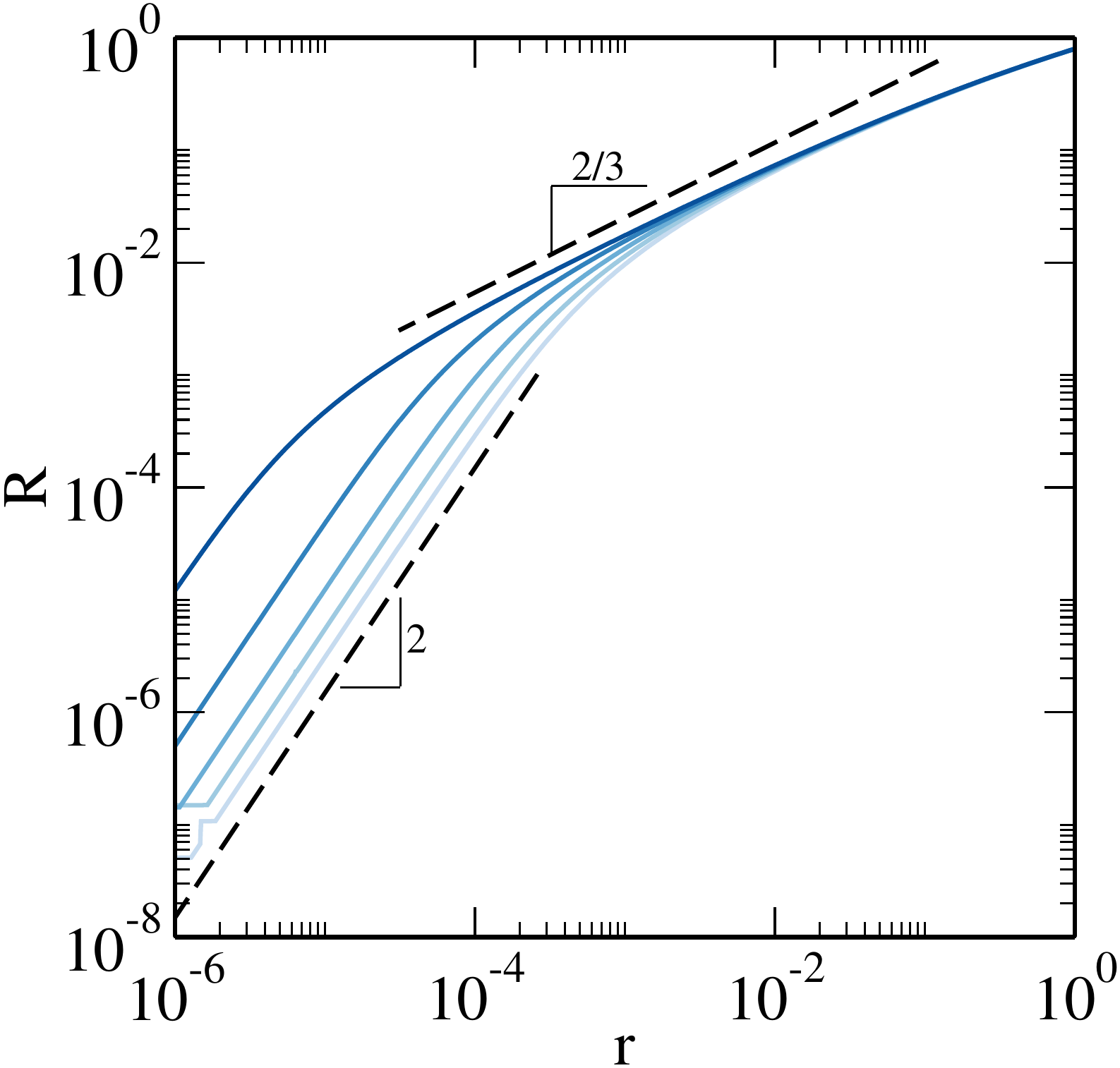}
\medskip
\centerline{(b)}
\end{minipage}\hfill
\begin{minipage}[t]{0.33\linewidth}\centering
\includegraphics[width=5cm]{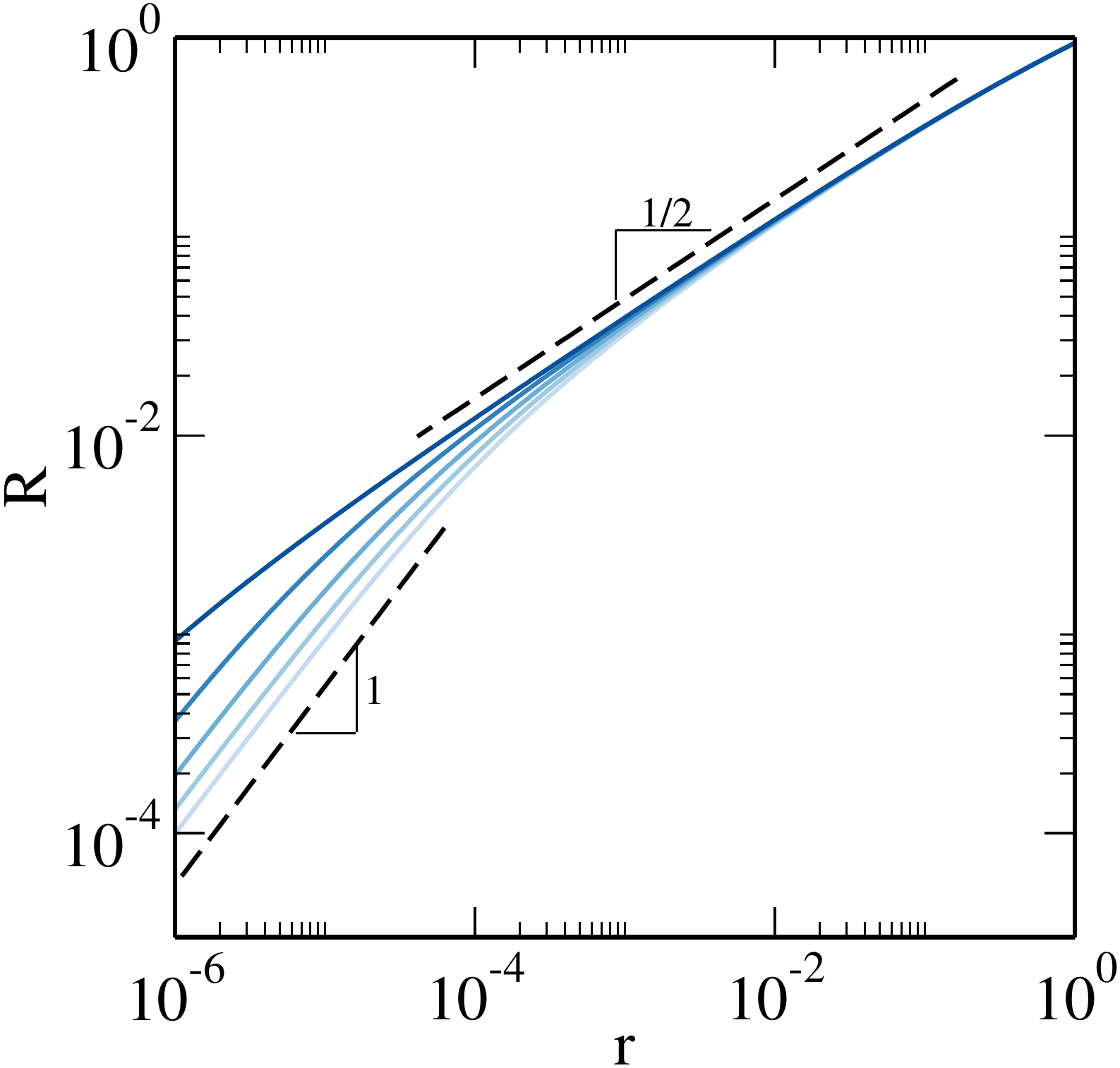}
\medskip
\centerline{(c)}
\end{minipage}
\caption{
Theoretical solutions of $R$ as a function of $r$ when $T^I\lesssim T^I_c$, i.e., $T^I=0.248,0.2485,0.249,0.2495,0.2499$ (from light blue to dark blue solid lines) with (a) $T^b=0.05$, (b) $T^b=0.1$, (c) $T^b=0.2$. Both internal links and bridge links are ER networks, with $\langle k \rangle=4$ and $\langle k^b \rangle=10$, respectively. Black dashed lines are drawn with the slope predicted by Table~\ref{tb_exponents} for different regimes.}
\label{fig_crossover}
\end{figure}

In Fig.~\ref{fig_crossover} we use different values of $T^I \lesssim T^I_c=1/(\kappa-1)$ to show the behavior of $R$ versus $r$ near the critical point, when $T^b$ is below, equal to, and above $T^b_c(r\rightarrow 0)$.
We can see a crossover, where the exponent $\epsilon$ is the same as the $\epsilon$ for $T^I = T^I_c$ when $r$ is not small enough (i.e., $r^*\ll r \ll 1$), and has the same $\epsilon$ as the asymptotic one for $T^I < T^I_c$ when $r$ is extremely small (i.e., $r\ll r^*$).
For example, when $T^b<T^b_c(r\rightarrow 0)=1/(\kappa^b-1)$, i.e., $T^b=0.05$, which is shown in Fig.~\ref{fig_crossover} (a), we can see the power-law behavior $R\propto r^{1/\epsilon}$ with $\epsilon=1$ (as in regime IV) when $r$ is relatively large,
but it is in the nonepidemic phase with $R=0$ (as in regime I) as $r\rightarrow 0$.
When $T^b=T^b_c(r\rightarrow 0)=1/(\kappa^b-1)$, i.e., $T^b=0.1$, which is shown in Fig.~\ref{fig_crossover} (b), the exponent $\epsilon$ changes from $3/2$ (as in regime V) to $1/2$ (as in regime II) as $r\rightarrow 0$.
When $T^b>T^b_c(r\rightarrow 0)$, e.g., $T^b=0.2$, which is shown in Fig.~\ref{fig_crossover} (c), the exponent $\epsilon$ changes from $2$ (as in regime VI) to $1$ (as in regime III) as $r\rightarrow 0$.

As was mentioned above, since the crossover occurs when $r^* R^{b*}\sim 1/s_{\max}$, and $s_{\max}$ depends on the internal transmissibility $T^I$, the values of $r^*$ and $R^*$ also depend on $T^I$. To be more explicit, they follow power laws of the difference between $T^I$ and its critical value, i.e., $r^*\sim|T^I-T^I_c|^{\beta_r}$, and $R^*\sim|T^I-T^I_c|^{\beta_R}$.
The values of $\beta_r$ and $\beta_R$ can be derived as the following.

Combined with the criteria $1/s_{\max} \sim r^*R^{b*}$, and considering the relation between $R^b$ and $R$ as in Eq.~(\ref{eq_rbvsr}), we will get
\begin{equation}
\label{eq_smaxvsrr}
1/s_{\max} \sim
\begin{cases}
r^*R^*, & \text{if } T^b<1/(\kappa^b-1) \\
r^*(R^*)^{\tau^b-2}, & \text{if } T^b=1/(\kappa^b-1) \\
r^*, & \text{if } T^b>1/(\kappa^b-1)
\end{cases}
.
\end{equation}
Since curves with different values of $T^I$ overlap for a relatively large $r$ (which is also verified in Fig.~\ref{fig_crossover}), we also have the relation $R^* \propto (r^*)^{1/\epsilon}$, where $\epsilon$ is the one for $T^I=1/(\kappa-1)$, respectively. If we combine $R^* \propto (r^*)^{1/\epsilon}$ with Eq.~(\ref{eq_smaxvsrr}), and knowing that $s_{\max}\sim |T^I-T^I_c|^{-1/\sigma}$, we obtain that
\begin{equation}
\label{eq_rstar}
r^* \sim
\begin{cases}
|T^I-T^I_c|^{(3-\tau)/\sigma} \sim |T^I-T^I_c|^\gamma, & \text{if } T^b<1/(\kappa^b-1) \\
|T^I-T^I_c|^{(1-(\tau-2)(\tau^b-2))/\sigma} \sim |T^I-T^I_c|^{(\tau^b-2)\gamma+(3-\tau^b)/\sigma}, & \text{if } T^b=1/(\kappa^b-1) \\
|T^I-T^I_c|^{1/\sigma}, & \text{if } T^b>1/(\kappa^b-1)
\end{cases}
.
\end{equation}
In any region, the crossover point $r^*$ goes to $0$ as $T^I$ approaches $T^I_c$, so we do not see a crossover unless $T^I$ is below but very close to $T^I_c$.

Knowing that the mean finite cluster size $\langle s \rangle\sim |T^I-T^I_c|^{-\gamma}$, and the largest finite cluster size $s_{\max}\sim |T^I-T^I_c|^{-1/\sigma}$,
Eq.~(\ref{eq_rstar}) can also be written as
\begin{equation}
\label{eq_rstarsimplified}
1/r^* \sim
\begin{cases}
\langle s \rangle, & \text{if } T^b<1/(\kappa^b-1) \\
\langle s \rangle^{\tau^b-2} {s_{\max}}^{3-\tau^b}, & \text{if } T^b=1/(\kappa^b-1) \\
s_{\max}, & \text{if } T^b>1/(\kappa^b-1)
\end{cases}
,
\end{equation}
and for all three regions of $T^b$, $R^* \sim |T^I-T^I_c|^{(\tau-2)/\sigma} \sim |T^I-T^I_c|^\beta$,
whose exponent is the same as the one in $R\propto |T-T_c|^\beta$ for an isolated network.

The scaling relation between $R$ and $r$ around the critical internal transmissibility ($T^I\lesssim T^I_c$) can then be written as
\begin{equation}
\label{eq_scaling}
R = R^*\,F\left(\frac{r}{r^*}\right),
\end{equation}
where $F(x)$ is given by $F(x)\sim x^{1/\epsilon}$ and
\begin{equation}
\label{eq_deltaprime}
\epsilon =
\begin{cases}
\epsilon (T^I<T^I_c), & \text{if } x \ll 1 \\
\epsilon (T^I=T^I_c), & \text{if } x \gg 1
\end{cases}
.
\end{equation}
(See Table \ref{tb_exponents} for values of $\epsilon$ for different values of $T^b$.)

\begin{figure}[h!]
\begin{minipage}[t]{0.33\linewidth}\centering
\includegraphics[width=5cm]{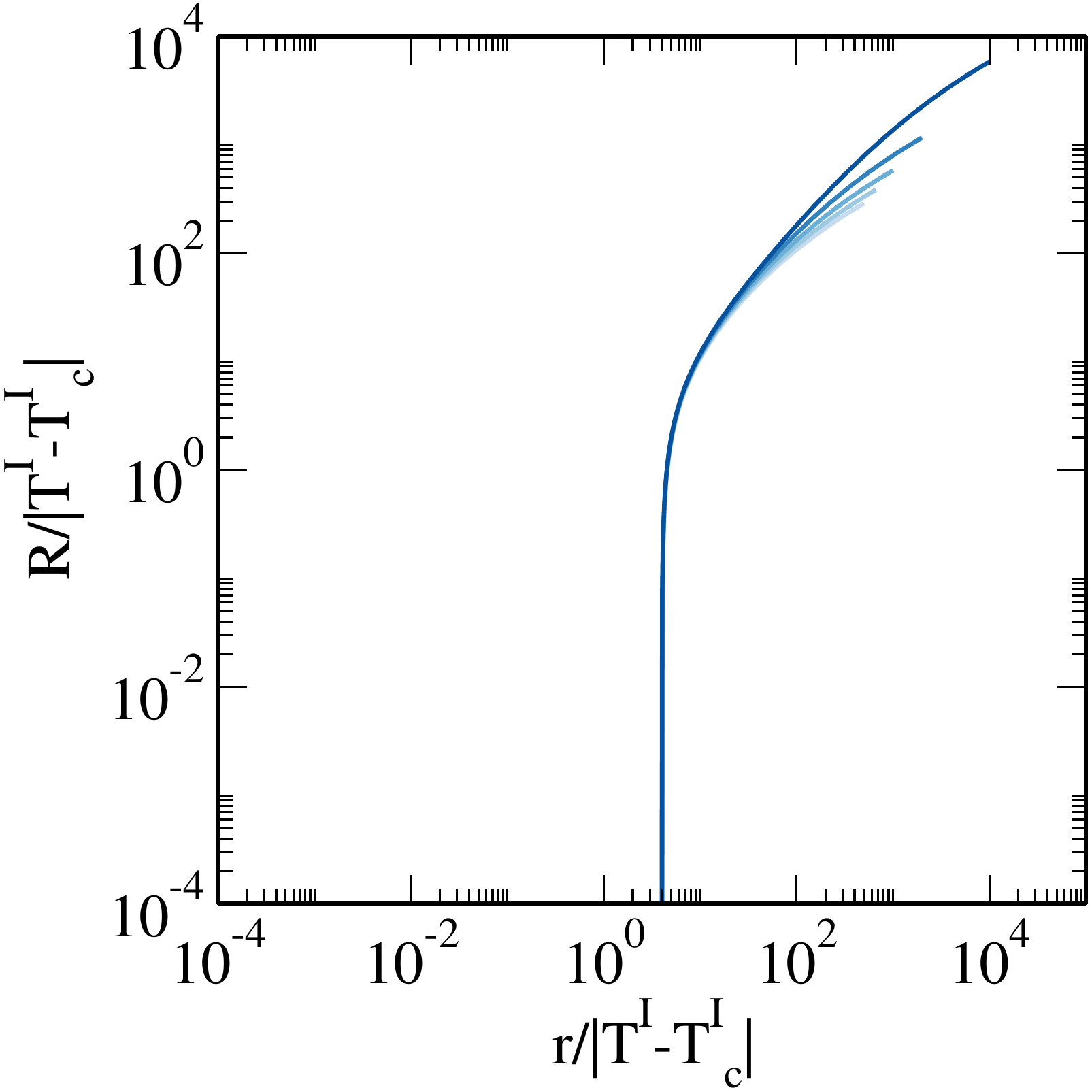}
\medskip
\centerline{(a)}
\end{minipage}\hfill
\begin{minipage}[t]{0.33\linewidth}\centering
\includegraphics[width=5cm]{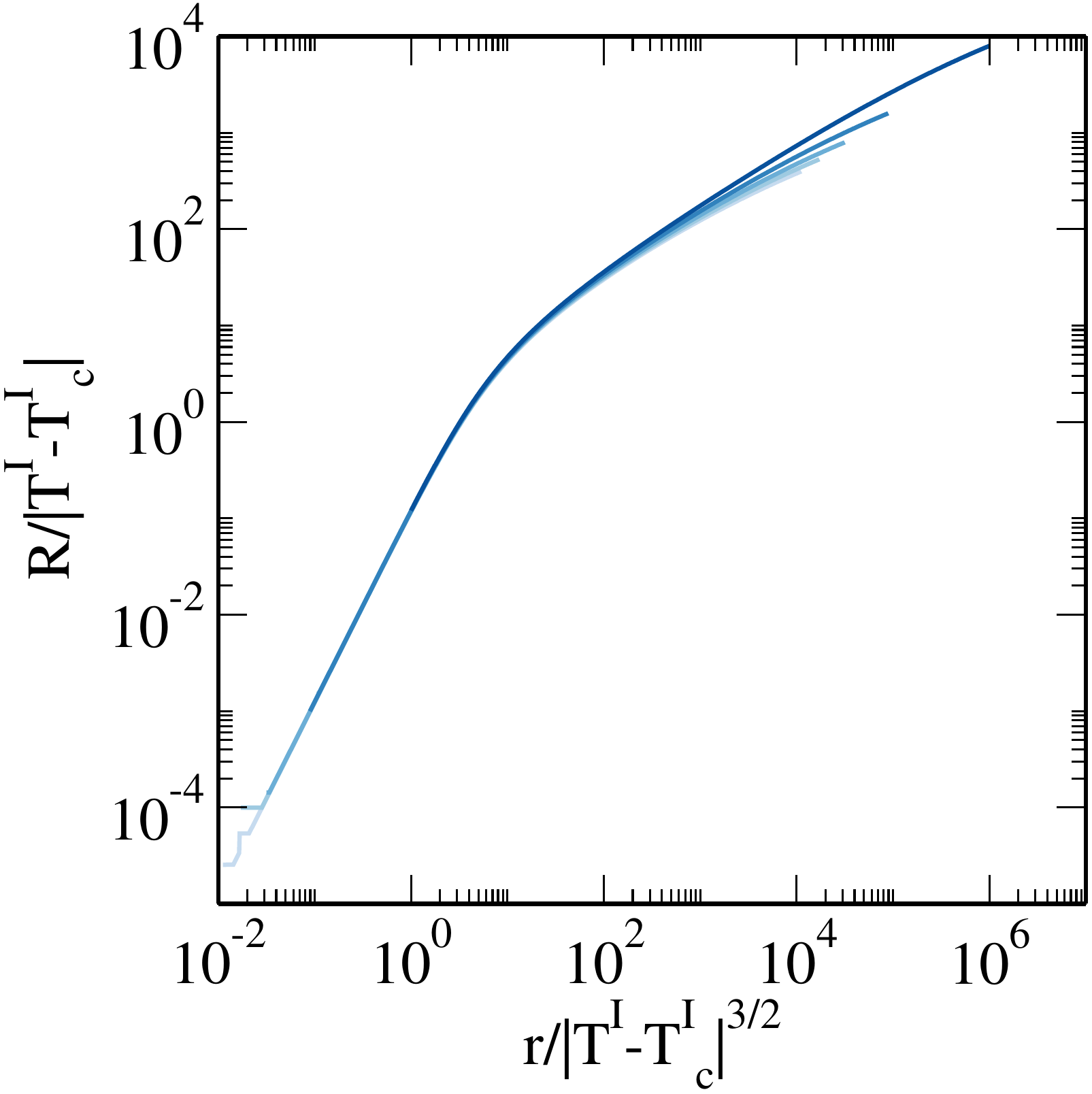}
\medskip
\centerline{(b)}
\end{minipage}\hfill
\begin{minipage}[t]{0.33\linewidth}\centering
\includegraphics[width=5cm]{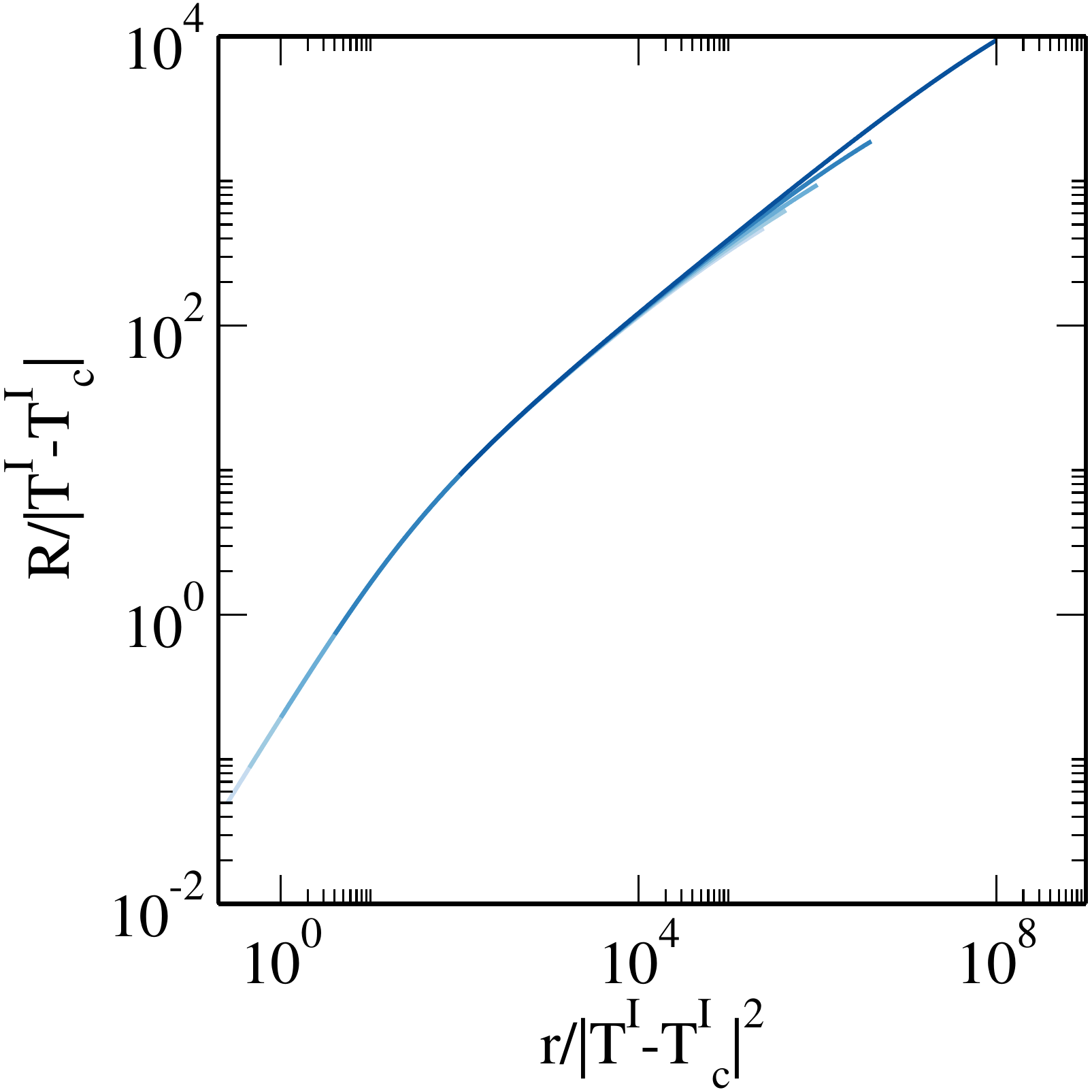}
\medskip
\centerline{(c)}
\end{minipage}
\caption{Theoretical solutions of $R/R^*$ as a function of $r/r^*$, where $R^*=|T^I-T^I_c|^{\beta_R}$ and $r^*=|T^I-T^I_c|^{\beta_r}$ when $T^I\lesssim T^I_c$, i.e., $T^I=0.248,0.2485,0.249,0.2495,0.2499$ (from light blue to dark blue solid lines) with (a) $T^b=0.05$, so that $\beta_r=1$ and $\beta_R=1$, (b) $T^b=0.1$, so that $\beta_r=3/2$ and $\beta_R=1$, and (c) $T^b=0.2$, so that $\beta_r=2$ and $\beta_R=1$. Both internal links and bridge links are ER networks, with $\langle k \rangle=4$ and $\langle k^b \rangle=10$, respectively. All curves with different $T^I\lesssim T^I_c$ collapse under the scaling relation.}
\label{fig_crossovercollapsed}
\end{figure}

In Fig.~\ref{fig_crossovercollapsed} we show the plot of $R$ versus $r$ rescaled by $R^*\sim|T^I-T^I_c|^{\beta_R}$ and $r^*\sim|T^I-T^I_c|^{\beta_r}$ for three different values of $T^b$.
Since both communities and the bridge links are all ER networks, $\tau=\tau^b=5/2$, $\gamma=1$, $\sigma=1/2$, and $\beta=1$.
For all values of $T^b$, $\beta_R=1$, which is the same as the exponent $\beta$ in $R \propto |T-T_c|^\beta$ of an isolated network, as mentioned above.
When $T^b<1/(\kappa^b-1)$, e.g., $T^b=0.05$, we have $\beta_r=\gamma=1$.
When $T^b=1/(\kappa^b-1)$, e.g., $T^b=0.1$, we have $\beta_r=(\tau^b-2)\gamma+(3-\tau^b)/\sigma=3/2$.
When $T^b>1/(\kappa^b-1)$, e.g., $T^b=0.2$, we have $\beta_r=1/\sigma=2$. We can see that the curves of $R/R^*$ versus $r/r^*$ for different values of $T^I \lesssim T^I_c$ collapse.

Empirically, there are cases when we can not or do not want to further reduce internal transmissibility, for example due to the shortage of facial masks or to avoid severe economic consequences, so that $T^I$ is below but close to its critical value.
In those scenarios, depending on the value of $T^I$, if there are too many open international airports, we may go through a section where shutting them down does not show a huge effect on the total number of infected individuals. However, as long as the internal transmissibility is below its critical value so that the disease can not spread massively within a community, if we keep reducing $r$, then after a point, which is the crossover, the total number of infected individuals is going to drop dramatically.
A good understanding of this crossover is going to help us better estimate the impact of epidemic strategies.


\section{Conclusions}

In this paper, we study the effect of bridge nodes to the final state of the SIR model, by mapping it to link percolation.
We find power-law asymptotic behaviors between $R$ and $r$ in different regimes, depending on how $T^I$ and $T^b$ are compared to their critical values.
The different exponents are related to the different mechanisms of how finite clusters in each community are connected into the GC of the whole system.
Additionally, around but below the critical point of internal transmissibility (when $T^I \lesssim T^I_c$), we find the crossover points $r^*$ such that $R$ versus $r$ follows a different power-law behavior when $r^* \ll r\ll 1$ compared to its asymptotic one (when $r\ll r^*$).
The methodology and results in this paper can easily be generalized for NONs with multiple communities.

The results can provide the authorities with helpful guidance on making decisions about epidemic strategies. They enable us to better anticipate the impacts of epidemic strategies before adopting them, and
help us understand why strategies like shutting down international airports had better be combined with adequate social distancing strategies to be more effective.


\section{Acknowledgments}

The authors would like to thank Shlomo Havlin for useful discussions.
J.M., L.D.V., and L.A.B. acknowledge support from NSF Grant No. PHY-1505000, DTRA Grant No. HDTRA-1-14-1-0017, and DTRA Grant No. HDTRA-1-19-1-0016. L.A.B. and L.D.V. thank UNMdP (Grant EXA956/20) and CONICET (Grant No. PIP 00443/2014) for financial support.



\begin{appendices}

\section{Physical and Non-Physical Critical Values}
\label{app_sec_critical}

\begin{figure}[htb]
\begin{minipage}[t]{0.49\linewidth}\centering
\includegraphics[width=5cm]{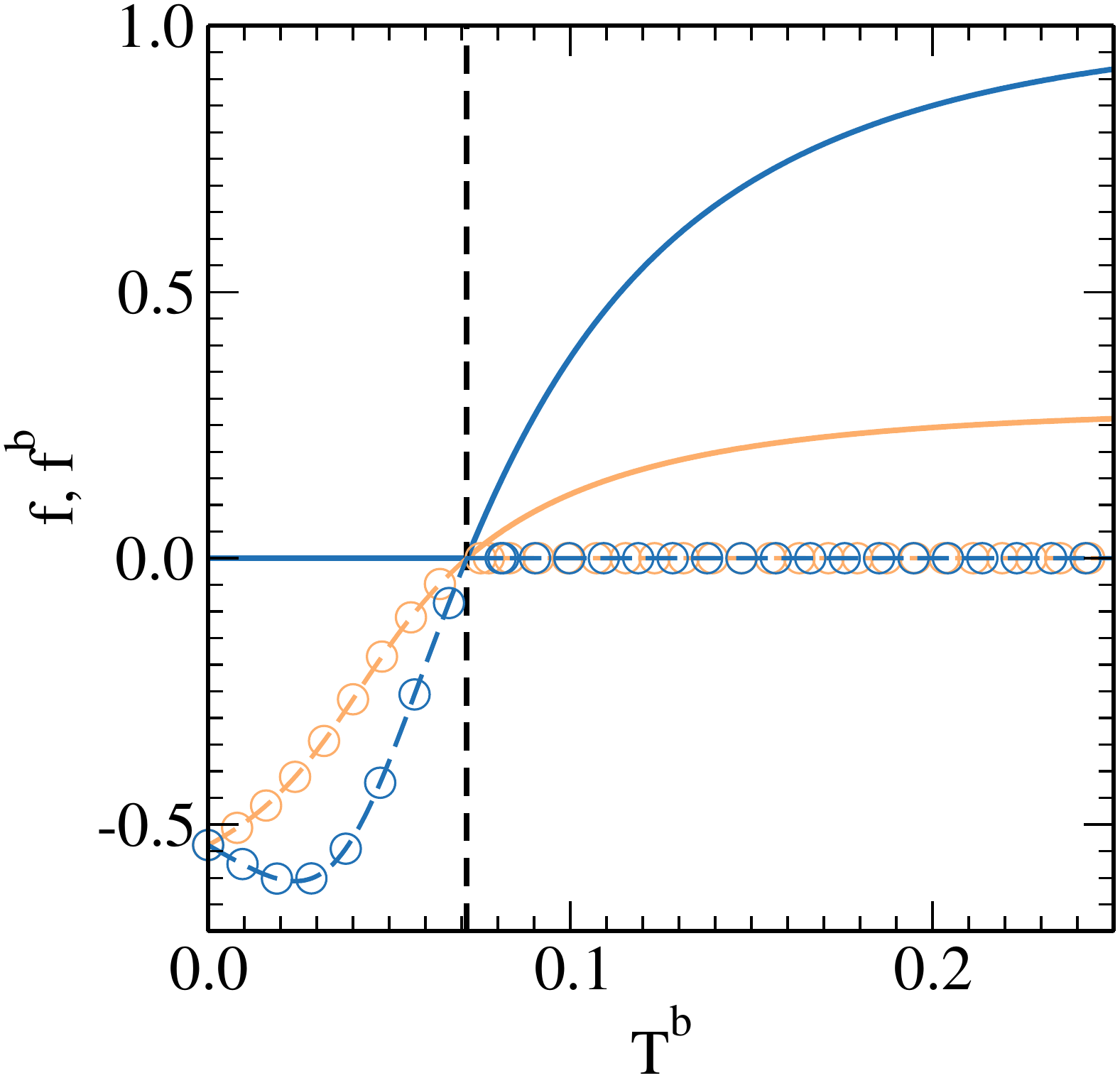}
\medskip
\centerline{(a)}
\end{minipage}\hfill
\begin{minipage}[t]{0.49\linewidth}\centering
\includegraphics[width=5cm]{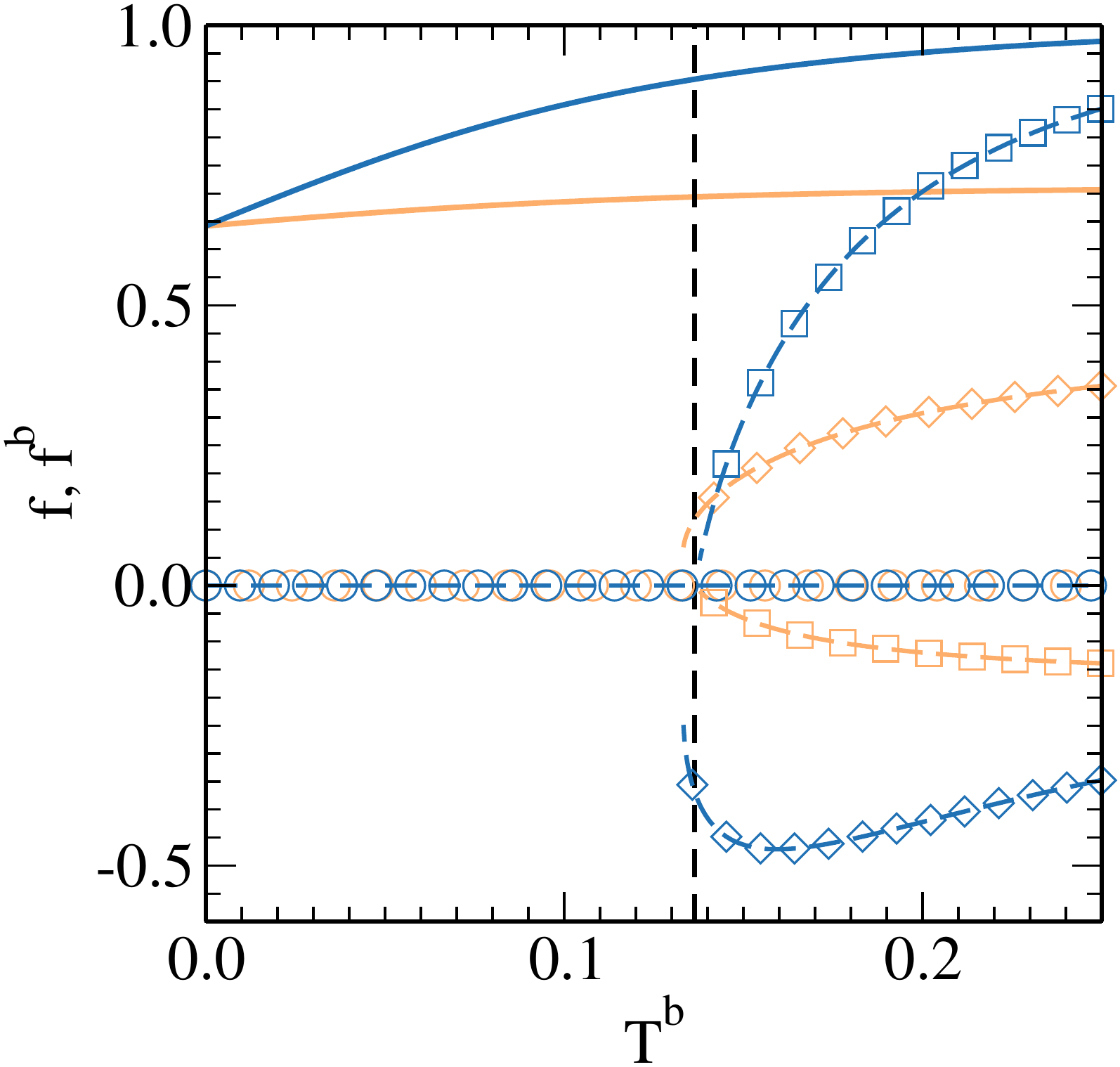}
\medskip
\centerline{(b)}
\end{minipage}
\caption{Numerical solutions of $f$ (orange) and $f^b$ (blue) of Eqs.~(\ref{eq_f}) and (\ref{eq_fb}) with $r=0.1$, given (a) $T^I=0.2$ and (b) $T^I=0.4$.
The only physical solution in each case is plotted in solid lines, and pairs of nonphysical solutions are in dashed lines with different symbols.
Both internal links and bridge links are ER networks, with $\langle k \rangle=4$ and $\langle k^b \rangle=10$, respectively.
The vertical dashed line represents the value of $T^b_c$ predicted by Eq.~(\ref{eq_tbc}).}
\label{app_fig_critical}
\end{figure}

When $T^I \le 1/(\kappa-1)$, e.g., $T^I=0.2$, as in Fig.~\ref{app_fig_critical} (a), Eq.~(\ref{eq_tbc}) gives the physical critical value $T^b_c$, from which point the physical solution of $f$ and $f^b$ becomes nontrivial. However, when $T^I > 1/(\kappa-1)$, e.g., $T^I=0.4$, as in Fig.~\ref{app_fig_critical} (b), Eq.~(\ref{eq_tbc}) gives the value of $T^b$ where more nonphysical solutions show up, while the physical solution stays smooth, so there is no critical phenomenon.

\section{Derivations of the Relations between $R$, $r$ and $R^b$ in Different Regimes}
\label{app_sec_exponents}

To derive how $R$ depends on $r R^b$, recall that $P(s) \sim s^{-\tau+1}\exp(-s/s_{\max})$, so Eq.~(\ref{eq_rsum}) becomes
\begin{equation}
\label{app_eq_r}
\begin{split}
R & = 1-\sum_{s=1}^\infty P(s)(1-r\,R^b)^s \\
    & \approx 1-\frac{\int_1^\infty s^{-\tau+1}e^{-s/s_{\max}}(1-r\,R^b)^s ds}{\int_1^\infty s^{-\tau+1}e^{-s/s_{\max}} ds} \\
    & = 1-\frac{\int_1^\infty s^{-\tau+1}e^{-s/s_{\max}}e^{s\ln(1-r\,R^b)} ds}{\int_1^\infty s^{-\tau+1}e^{-s/s_{\max}} ds} \\
    & \approx 1-\frac{\int_1^\infty s^{-\tau+1}e^{-s(1/s_{\max}+r\,R^b)} ds}{\int_1^\infty s^{-\tau+1}\exp(-s/s_{\max}) ds}.
\end{split}
\end{equation}

When $T^I=T^I_c=1/(\kappa-1)$, so $s_{\max}$ diverges and thus $P(s) \sim s^{-\tau+1}$,
or if $T^I\lesssim T^I_c=1/(\kappa-1)$, but $r$ is not too small,
so that $1/s_{\max}\ll r R^b \ll 1$ and thus $1/s_{\max}$ can be ignored,
\begin{equation}
\label{app_eq_rat}
\begin{split}
R & \approx 1-\frac{\int_1^\infty s^{-\tau+1}e^{-r\,R^b\,s} ds}{\int_1^\infty s^{-\tau+1} ds} \\
    & = 1-(\tau-2) \int_1^\infty s^{-\tau+1}e^{-r\,R^b\,s} ds \\
    & = 1-(\tau-2)(r\,R^b)^{\tau-2} \int_{r\,R^b}^\infty u^{-\tau+1}e^{-u} du, \text{ where }u=r\,R^bs \\
    & = 1-(\tau-2)(rR^b)^{\tau-2} \left[\frac{(rR^b)^{-\tau+2}}{\tau-2}e^{-rR^b} - \int_{rR^b}^\infty \frac{u^{-\tau+2}}{\tau-2}e^{-u} du \right] \\
    & \approx (r\,R^b)^{\tau-2} \int_{r\,R^b}^\infty u^{-\tau+2}e^{-u} du \\
    & \propto (r\,R^b)^{\tau-2}.
\end{split}
\end{equation}

When $T^I<T^I_c=1/(\kappa-1)$, so $s_{\max}$ is finite and thus $P(s) \sim s^{-\tau+1}\exp(-s/s_{\max})$,
or if $T^I\lesssim T^I_c=1/(\kappa-1)$, but $r$ is very small,
so that $1/s_{\max}$ can not be ignored compared with $r R^b$,
\begin{equation}
\label{app_eq_rbelow}
\begin{split}
R & \approx 1-\frac{\int_1^\infty s^{-\tau+1}e^{-s/s_{\max}}(1-r\,R^b)^s ds}{\int_1^\infty s^{-\tau+1}e^{-s/s_{\max}} ds} \\
    & \approx 1-\frac{\int_1^\infty s^{-\tau+1}e^{-s/s_{\max}}(1-r\,R^bs) ds}{\int_1^\infty s^{-\tau+1}e^{-s/s_{\max}} ds} \\
    & = r\,R^b \frac{\int_1^\infty s^{-\tau+2}e^{-s/s_{\max}} ds}{\int_1^\infty s^{-\tau+1}e^{-s/s_{\max}} ds} \\
    & \propto r\,R^b.
\end{split}
\end{equation}

When $T^I>T^I_c=1/(\kappa-1)$, it is in the epidemic phase and
$R$ does not approach $0$ as $r\rightarrow 0$,
so there is no power-law relation between $R$ and $r R^b$ in this regime.

On the other hand, to derive how $R^b$ depends on $R$, recall that $P^b(s) \sim s^{-\tau^b+1}\exp(-s/s^b_{\max})$.
When $T^b=T^b_c(r\rightarrow 0)=1/(\kappa^b-1)$,
so $s^b_{\max}$ diverges and $P^b(s) \sim s^{-\tau^b+1}$,
\begin{equation}
\label{app_eq_rbat}
\begin{split}
R^b & \approx 1-\sum_{s=1}^\infty P^b(s)(1-R)^s \\
    & \approx 1-\frac{\int_1^\infty s^{-\tau^b+1}(1-R)^s ds}{\int_1^\infty s^{-\tau^b+1} ds} \\
    & = 1-(\tau^b-2) \int_1^\infty s^{-\tau^b+1}e^{s\ln(1-R)} ds \\
    & \approx 1-(\tau^b-2) \int_1^\infty s^{-\tau^b+1}e^{-Rs} ds \\
    & = 1-(\tau^b-2)R^{\tau^b-2} \int_R^\infty u^{-\tau^b+1}e^{-u} du, \text{ where }u=Rs \\
    & = 1-(\tau^b-2)R^{\tau^b-2} \left[\frac{R^{-\tau^b+2}}{\tau^b-2}e^{-R} - \int_R^\infty \frac{u^{-\tau^b+2}}{\tau^b-2}e^{-u} du \right] \\
    & \approx R^{\tau^b-2} \int_R^\infty u^{-\tau^b+2}e^{-u} du \\
    & \propto R^{\tau^b-2}.
\end{split}
\end{equation}

When $T^b<T^b_c(r\rightarrow 0)=1/(\kappa^b-1)$, and thus $s^b_{\max}<\infty$,
\begin{equation}
\label{app_eq_rbbelow}
\begin{split}
R^b & \approx 1-\sum_{s=1}^\infty P^b(s)(1-R)^s \\
    & \approx 1-\frac{\int_1^\infty s^{-\tau^b+1}e^{-s/s^b_{\max}}(1-R)^s ds}{\int_1^\infty s^{-\tau^b+1}e^{-s/s^b_{\max}} ds} \\
    & \approx 1-\frac{\int_1^\infty s^{-\tau^b+1}e^{-s/s^b_{\max}}(1-Rs) ds}{\int_1^\infty s^{-\tau^b+1}e^{-s/s^b_{\max}} ds} \\
    & = R \frac{\int_1^\infty s^{-\tau^b+2}e^{-s/s^b_{\max}} ds}{\int_1^\infty s^{-\tau^b+1}e^{-s/s^b_{\max}} ds} \\
    & \propto R.
\end{split}
\end{equation}

When $T^b>T^b_c(r\rightarrow 0)=1/(\kappa^b-1)$,
most bridge nodes are connected into one big cluster through bridge links, so Eqs.~(\ref{eq_rbsum})~and~(\ref{eq_rbsumapprox}) do not hold and
$R^b$ is not a power law of $R$.


\section{Extension: Regimes when $P^A(k)\ne P^B(k)$}
\label{app_sec_panepb}

When the two communities of the system have different degree distributions,
i.e., $P^A(k)\ne P^B(k)$,
a similar methodology can be applied to make predictions about the asymptotic power-law relations between $R$ and $r$ in different regimes.

In this case,
instead of Eqs.~(\ref{eq_f}) and (\ref{eq_fb}),
we are going to have the following theoretical equations:
\begin{eqnarray}
\label{app_eq_fa}
f^A &=& (1-r)\left[1-G_1^A(1-T^I f^A)\right] + r\left[1-G_1^A(1-T^I f^A)G_0^b(1-T^bf^{A,b})\right],\\
\label{app_eq_fb}
f^B &=& (1-r)\left[1-G_1^B(1-T^I f^B)\right] + r\left[1-G_1^B(1-T^I f^B)G_0^b(1-T^bf^{B,b})\right],\\
\label{app_eq_fab}
f^{A,b} &=& 1-G_1^b(1-T^bf^{B,b})G_0^B(1-T^I f^B),\\
\label{app_eq_fbb}
f^{B,b} &=& 1-G_1^b(1-T^bf^{A,b})G_0^A(1-T^I f^A),
\end{eqnarray}
where $f^A$ (or $f^B$) is the probability to expand a branch to the infinity through an internal link in community A (or B), $f^{A,b}$ (or $f^{B,b}$) is the probability to expand a branch to the infinity through a bridge link,
which starts from a bridge node in community A (or B);
instead of Eqs.~(\ref{eq_r}) and (\ref{eq_rb}),
we are going to have
\begin{eqnarray}
\label{app_eq_ra} R^A & = &(1-r)\left[1-G_0^A(1-T^I f^A)\right] + r\left[1-G_0^A(1-T^I f^A)G_0^b(1-T^bf^{A,b})\right], \\
\label{app_eq_rb} R^B & = &(1-r)\left[1-G_0^B(1-T^I f^B)\right] + r\left[1-G_0^B(1-T^I f^B)G_0^b(1-T^bf^{B,b})\right], \\
\label{app_eq_r_avg}
R & = & (R^A+R^B)/2, \\
\label{app_eq_rab}
R^{A,b} &=& 1-G_0^b(1-T^bf^{A,b})G_0^A(1-T^I f^A), \\
\label{app_eq_rbb}
R^{B,b} &=& 1-G_0^b(1-T^bf^{B,b})G_0^B(1-T^I f^B), \\
\label{app_eq_rb_avg}
R^b & = & (R^{A,b}+R^{B,b})/2.
\end{eqnarray}
It is easy to see that the equations above will be reduced to Eqs.~(\ref{eq_f})-(\ref{eq_rb}) if $P^A(k)=P^B(k)=P(k)$.

To predict the relation between $R^A$, $R^B$, and $R$ as a function of $r$,
we assume $\kappa^A<\kappa^B$,
without loss of generality.
Then there will be 15 regimes,
namely, the five regimes for $T^I$ ($T^I<\frac{1}{\kappa^B-1}<\frac{1}{\kappa^A-1},T^I=\frac{1}{\kappa^B-1}<\frac{1}{\kappa^A-1},\frac{1}{\kappa^B-1}<T^I<\frac{1}{\kappa^A-1},\frac{1}{\kappa^B-1}<T^I=\frac{1}{\kappa^A-1},\frac{1}{\kappa^B-1}<\frac{1}{\kappa^A-1}<T^I$),
combined with the three regimes for $T^b$ ($T^b<\frac{1}{\kappa^b-1},T^b=\frac{1}{\kappa^b-1},T^b>\frac{1}{\kappa^b-1}$).

Here, we select two regimes as examples: (a) $T^I<\frac{1}{\kappa^B-1}<\frac{1}{\kappa^A-1},T^b=\frac{1}{\kappa^b-1}$, and (b) $T^I=\frac{1}{\kappa^B-1}<\frac{1}{\kappa^A-1},T^b=\frac{1}{\kappa^b-1}$.

In case (a),
where $T^I<\frac{1}{\kappa^A-1}$ and $T^I<\frac{1}{\kappa^B-1}$,
using a similar methodology as in Eq.~(\ref{eq_rvsrrb}),
we will get 
$R^A \propto rR^{A,b}$ and $R^B \propto rR^{B,b}$.
Since $T^b=\frac{1}{\kappa^b-1}$,
using a similar methodology as in Eq.~(\ref{eq_rbvsr}),
but substituting $(1-R)$ by $\sqrt{(1-R^A)(1-R^B)}$
\footnote[2]{For a finite cluster of bridge nodes, especially if it is large enough to belong to the GC, there are approximately the same number of nodes that belong to each community, so instead of $R^b\approx 1-\sum_s^\infty P^b(s)(1-R)^s$ as in Eq.~(\ref{eq_rbsumapprox}), we will have $R^b\approx 1-\sum_s^\infty P^b(s)(1-R^A)^{s/2}(1-R^B)^{s/2}$ in this case.},
we will get $R^{A,b}\sim R^{B,b}\sim R^b\propto \left[1-\sqrt{(1-R^A)(1-R^B)}\right]^{\tau^b-2}$.
In the case of $r\rightarrow 0$ so that $R^A,R^B\rightarrow 0$,
as well as $\kappa^A<\kappa^B$ so that $R^B\gg R^A$,
we will have $1-\sqrt{(1-R^A)(1-R^B)}\approx \frac{R^A+R^B}{2}=R\propto R^B$.
That is to say,
$R$ is dominated by the community with a larger $\kappa$.
As a result,
we will have $R^b\propto (R^B)^{\tau^b-2}$.
Combining $R^A \propto rR^b,R^B \propto rR^b$ with $R^b\propto (R^B)^{\tau^b-2}$,
we will get
$R^A\propto r^{1/\epsilon^A}$,
where $\epsilon^A=1-(\tau^b-2)$, and $R^B\propto r^{1/\epsilon^B}$,
where $\epsilon^B=1-(\tau^b-2)$.
We have $\epsilon^A=\epsilon^B$,
as expected,
since $T^I$ is below critical in both communities,
and the $\epsilon$ as in $R\propto r^{1/\epsilon}$ has the same value $\epsilon=1-(\tau^b-2)$ as well.

In case (b),
where $T^I<\frac{1}{\kappa^A-1}$ and $T^I=\frac{1}{\kappa^B-1}$,
similarly,
we will get 
$R^A \propto rR^b$ and $R^B \propto (rR^b)^{\tau-2}$.
Since $T^b=\frac{1}{\kappa^b-1}$,
we still get $R^b\propto \left[1-\sqrt{(1-R^A)(1-R^B)}\right]^{\tau^b-2}\propto \left(\frac{R^A+R^B}{2}\right)^{\tau^b-2}\propto (R^B)^{\tau^b-2}$.
Combining them,
we will get
$R^A\propto r^{1/\epsilon^A}$,
where $\epsilon^A=1-(\tau-2)(\tau^b-2)$, and $R^B\propto r^{1/\epsilon^B}$,
where $\epsilon^B=\frac{1}{\tau-2}-(\tau^b-2)$.
The $\epsilon$ as in $R\propto r^{1/\epsilon}$ is dominated by $\epsilon^B$,
i.e., $\epsilon=\epsilon^B=\frac{1}{\tau-2}-(\tau^b-2)$.

\begin{figure}[t] \begin{minipage}[t]{0.45\linewidth}\centering
\includegraphics[width=6cm]{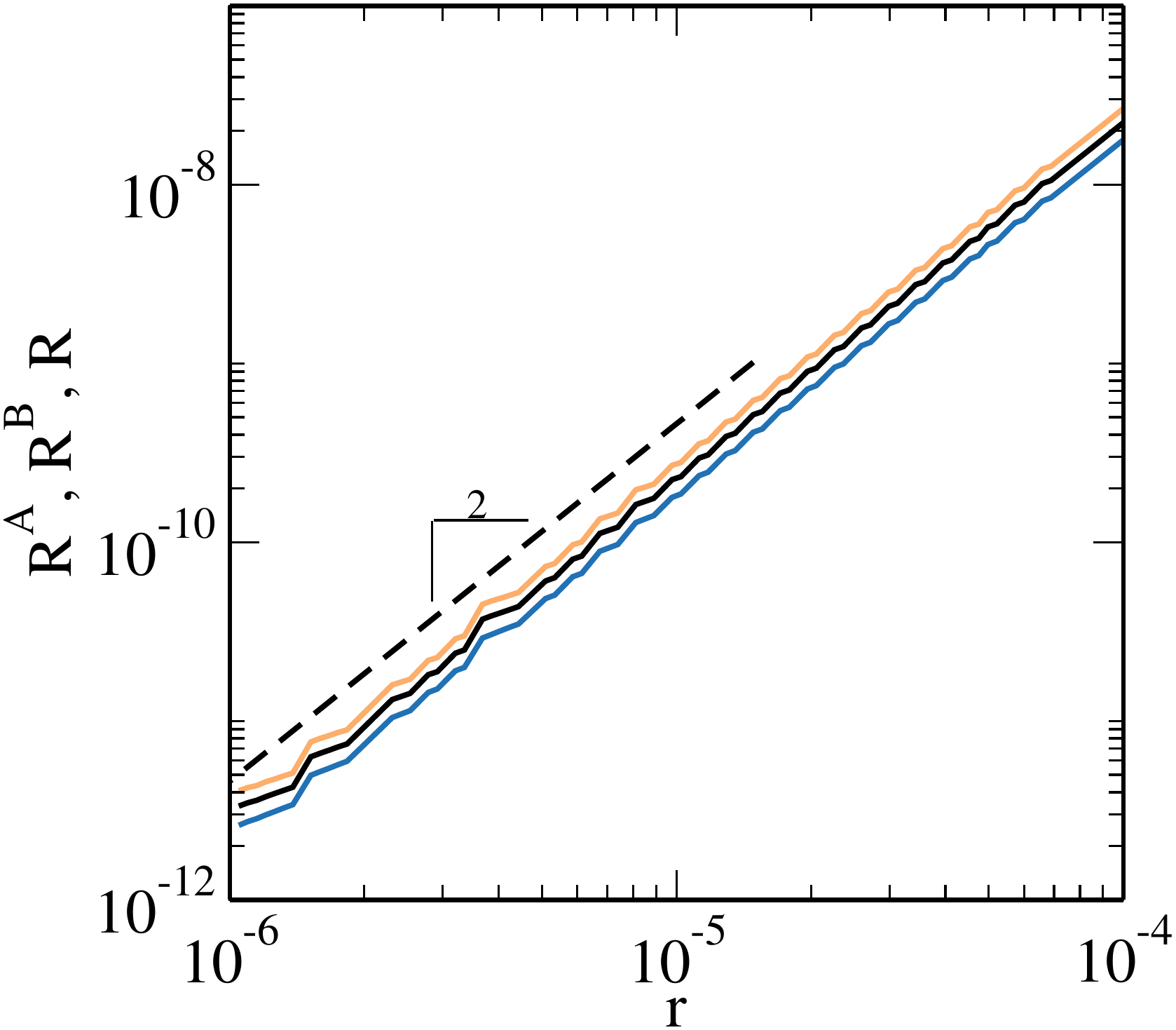}
\medskip
\centerline{(a)}
\end{minipage}\hfill
\begin{minipage}[t]{0.45\linewidth}\centering
\includegraphics[width=6cm]{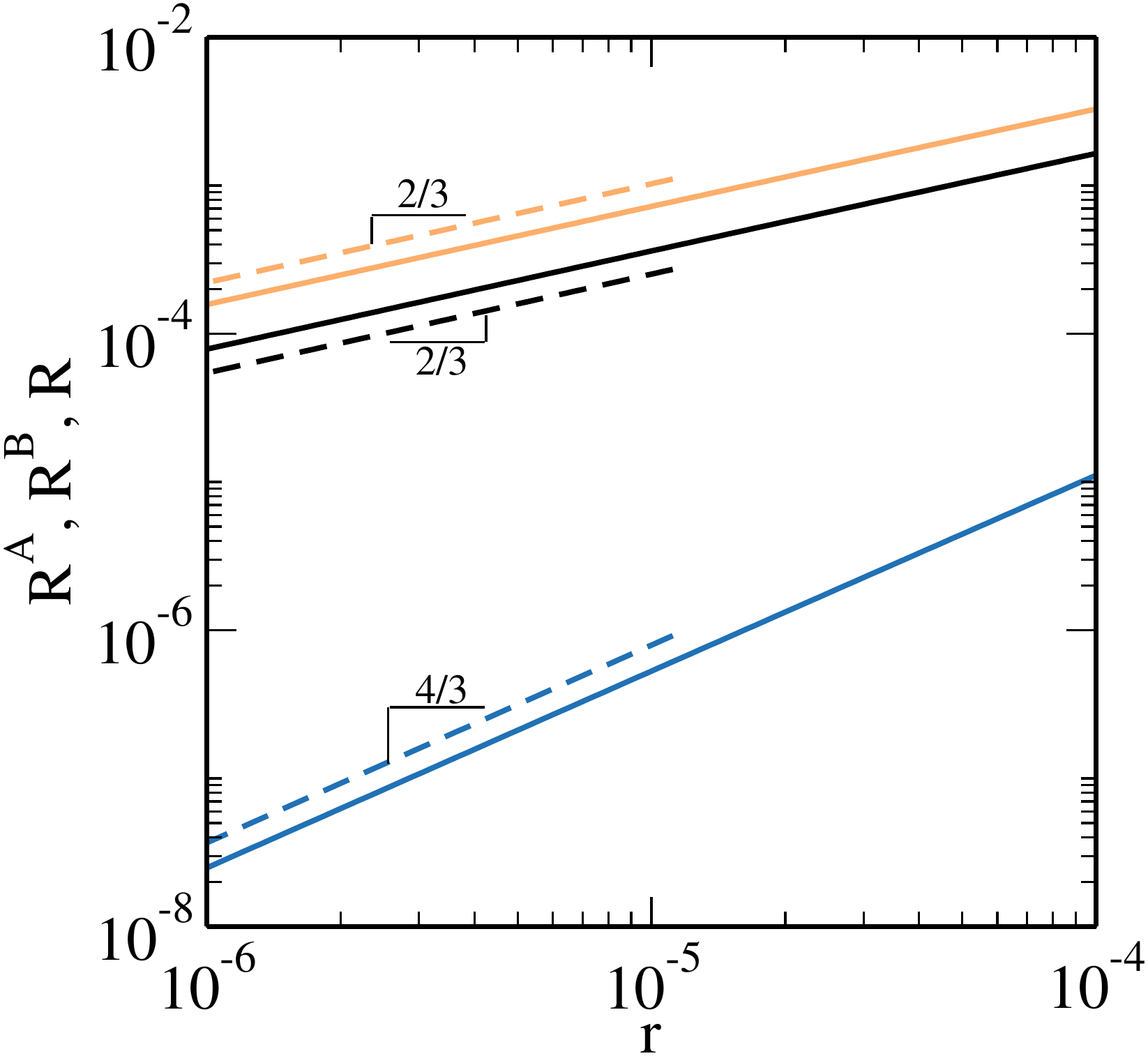}
\medskip
\centerline{(b)}
\end{minipage}
\caption{$R^A$ (blue), $R^B$ (orange), and $R=(R^A+R^B)/2$ (black) as a function of $r$, for two example regimes when $P^A(k)\ne P^B(k)$: (a) $T^I=0.0625$, $T^b=0.1$; (b) $T^I=0.125$, $T^b=0.1$. Both internal links and bridge links are ER networks, with $\langle k^A\rangle=4$, $\langle k^B\rangle=8$,
and $\langle k^b\rangle=10$.
In each regime, numerical solutions of Eqs.~(\ref{app_eq_fa})-(\ref{app_eq_r_avg}) are plotted in solid lines, and dashed lines are drawn with predicted slopes.}
\label{app_fig_exponents_ab}
\end{figure}

As in Fig.~\ref{app_fig_exponents_ab},
we consider a system where both internal and external links are ER networks,
with $\langle k^A\rangle=4$,
$\langle k^B\rangle=8$,
and $\langle k^b\rangle=10$.
As a result,
$\tau^I=\tau^b=5/2$,
$\kappa^A=5$ such that $T^A_c=\frac{1}{\kappa^A-1}=0.25$,
$\kappa^B=9$ such that $T^B_c=\frac{1}{\kappa^B-1}=0.125$,
and $\kappa^b=11$ such that $T^b_c(r\rightarrow 0)=\frac{1}{\kappa^b-1}=0.1$.
We can see from Fig.~\ref{app_fig_exponents_ab}
that the numerical solutions of $R^A,R^B$, and $R$ from Eqs.~(\ref{app_eq_fa})-(\ref{app_eq_r_avg}) (solid lines) agree well with dashed lines, whose slopes are predicted as above, as $r\rightarrow 0$.
It can be verified that a similar methodology can be used to give correct predictions for all 15 regimes.


\section{Simulation Results Compared with Numerical Solutions}
\label{app_sec_compare}

\begin{figure}[t]
\begin{minipage}[t]{0.49\linewidth}\centering
\includegraphics[width=5cm]{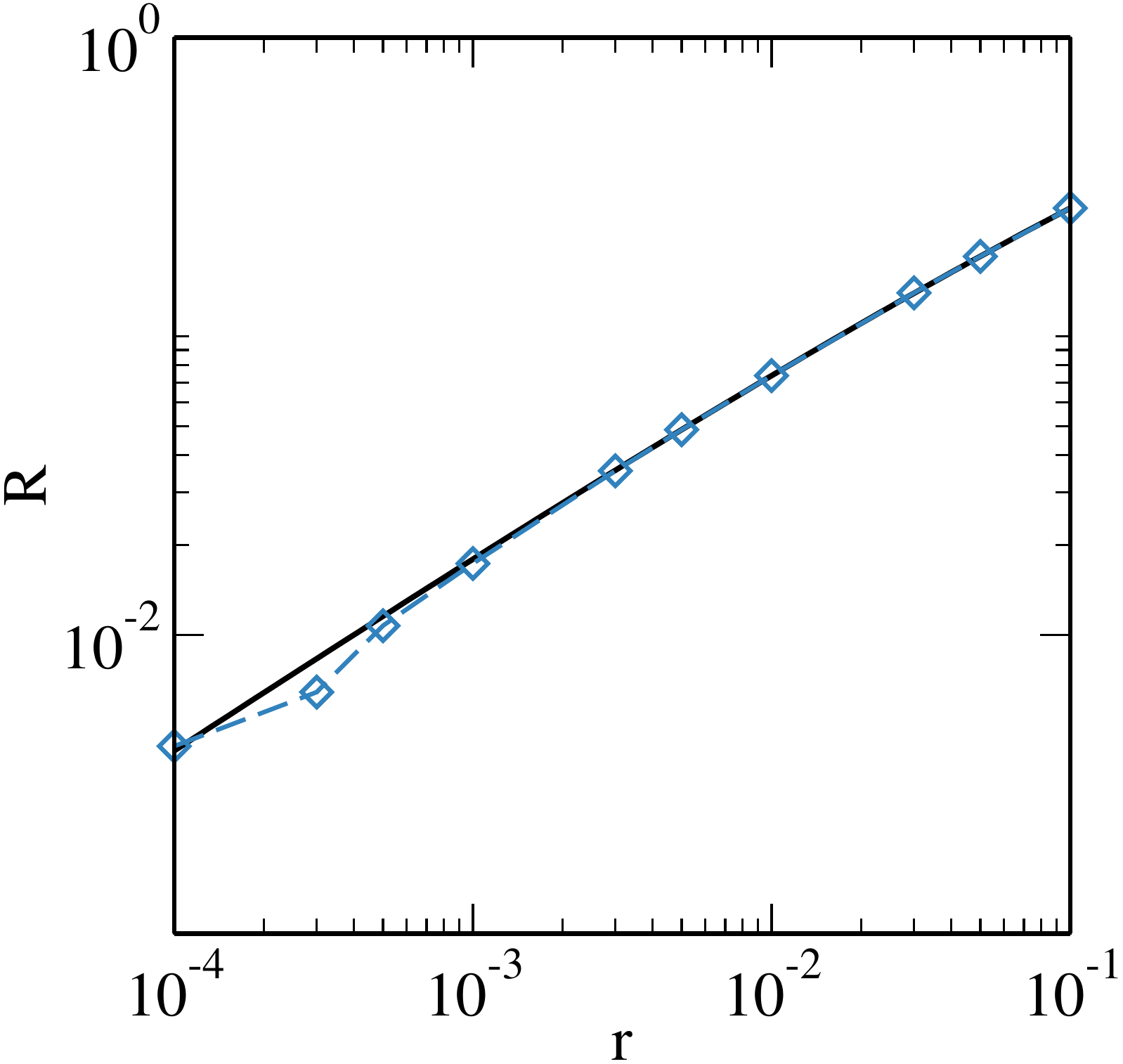}
\medskip
\centerline{(a)}
\end{minipage}\hfill
\begin{minipage}[t]{0.49\linewidth}\centering
\includegraphics[width=5cm]{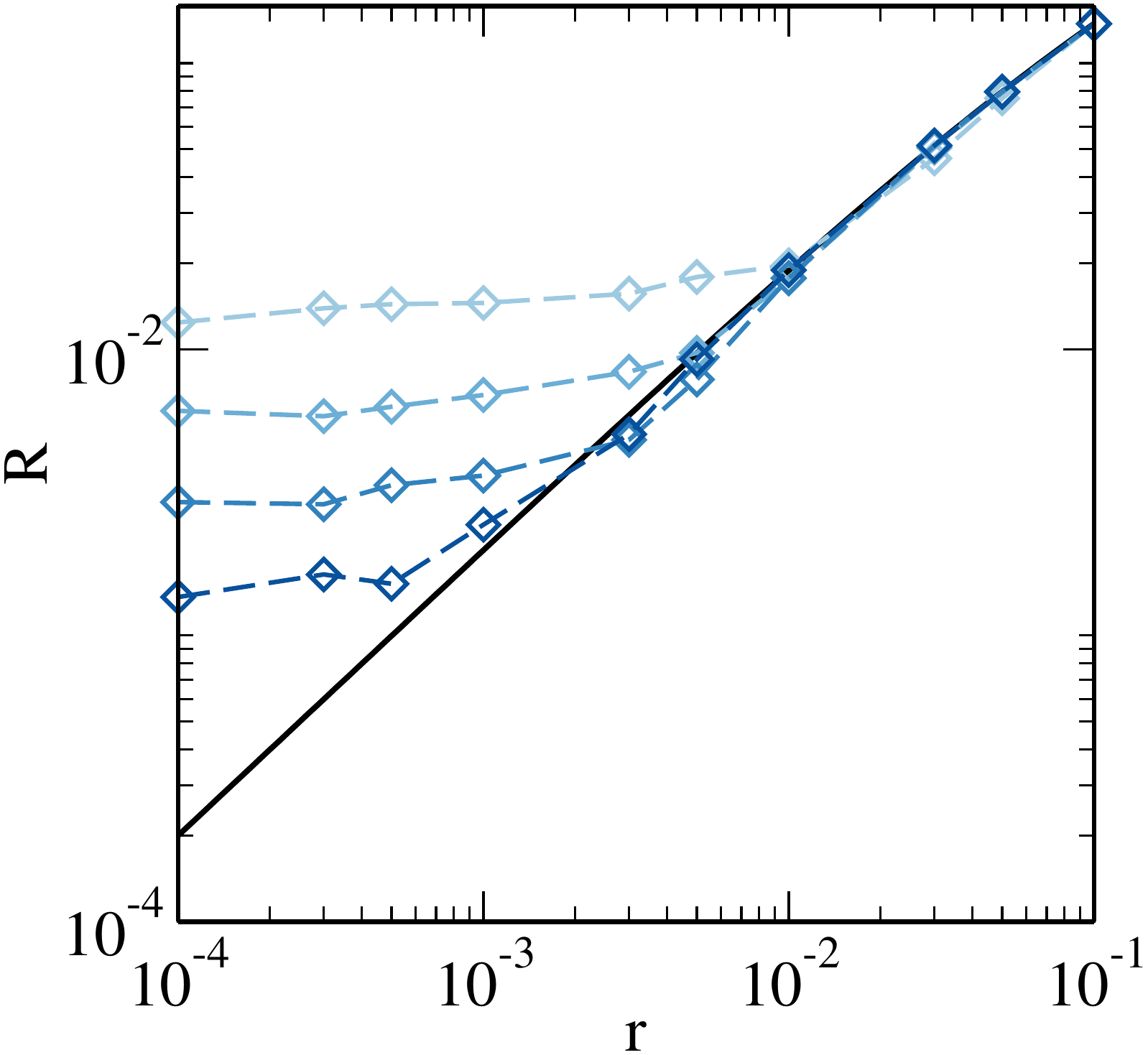}
\medskip
\centerline{(b)}
\end{minipage}
\caption{
Theoretical solutions (black solid lines) compared with simulation results (from link percolation mapping, dashed lines with diamond symbols) with (a) $T^I=0.25$, $T^b=0.1$, and system size $N_A=N_B=10^7$ (blue), and (b) $T^I=0.25$, $T^b=0.05$, with system sizes $N_A=N_B=10^5,10^6,10^7,10^8$ (from light blue to dark blue). Both internal links and bridge links are ER networks, with $\langle k \rangle=4$ and $\langle k^b \rangle=10$, respectively. For the simulations, $k_{\min}=0$, $k_{\max}=100$, and are averaged over $100$ realizations.}
\label{app_fig_rvsrmultiple}
\end{figure}

\begin{figure}[htb]
\centering
\includegraphics[width=10cm]{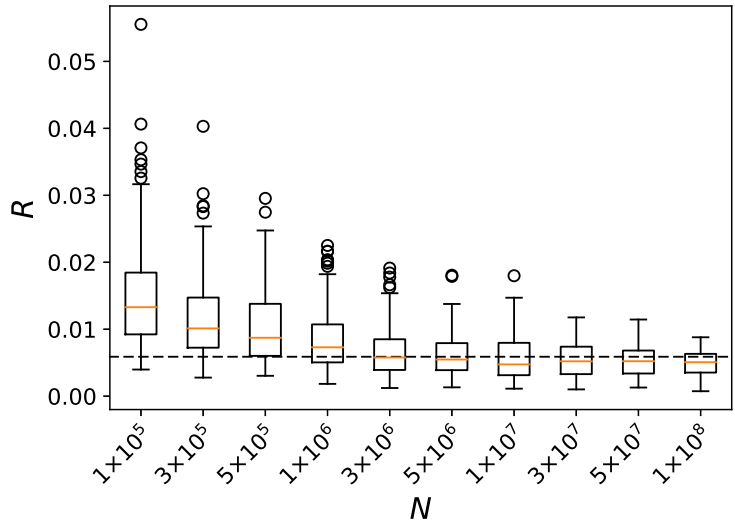}
\caption{Box plots of simulation results (from link percolation mapping) compared with the theoretical solution (dashed horizontal line), for different system sizes with $r=0.003$, $T^I=0.25$, and $T^b=0.05$.
Both internal and bridge links are ER networks, with $\langle k \rangle=4$, and $\langle k^b \rangle=10$, respectively.
For the simulations, $k_{\min}=0$, $k_{\max}=100$, and are plotted with $200$ realizations.}
\label{app_fig_boxplot}
\end{figure}

In Fig.~\ref{app_fig_rvsrmultiple},
we show the simulation results of the link percolation mapping and numerical solutions of Eqs.~(\ref{eq_f})-(\ref{eq_r}) when both internal links and bridge links are ER networks, with $\langle k \rangle=4$, and $\langle k^b \rangle=10$. When $T^I=1/\langle k \rangle=0.25$, $T^b=1/\langle k^b \rangle=0.1$, the simulation agrees well with theoretical solutions [see Fig.~\ref{app_fig_rvsrmultiple} (a)]. When $T^I=0.25$, $T^b=0.05$, a finite-size effect shows up and a much larger system size is required in order to obtain the theoretical results. From Fig.~\ref{app_fig_rvsrmultiple} (b) we can see that as the system size increases, the simulation results converge to the theoretical solution.
This is further verified in Fig.~\ref{app_fig_boxplot}, in which we show the box plots of the simulation results of $R$ for different system sizes. We can see that as system size increases, the distribution of $R$ narrows and converges to the theoretical solution (horizontal dashed line).

\end{appendices}

\bibliography{main}

\end{document}